# Table of Contents

# A Self-Consistent Two-Carrier Drift–Diffusion Framework for p–n Junctions: Extended Diode Equation, Power Dissipation, and Base Transit Time

Jianhong Yang*, Tinglan Li, and Gonghao Sun

*School of Physical Science and Technology, Lanzhou University, Lanzhou 730000, China*

*Corresponding author: yangjh@lzu.edu.cn

**Abstract.** The classical theory of p–n junctions derives the current–voltage characteristic from minority-carrier diffusion alone: the majority carriers are an inert background, the quasi-neutral base is field-free, and the junction voltage is identified with the externally applied bias. These premises fail together where device performance matters most, namely under high bias and high-level injection, and the theory's high-injection current roll-off is reproduced only by appending an empirical series resistance that lies outside the physics of the junction. Here we develop a self-consistent two-carrier drift–diffusion framework that removes these limitations without introducing any assumption beyond the classical ones. Excess majority carriers are shown to be electrostatically forced transport participants, and the ambipolar diffusion equation, closed by exact two-sided boundary conditions, yields a unified excess-carrier profile. The self-consistent base electric field separates into an ohmic and a Dember component, and all four current components follow in closed form, their sum being position-independent. The framework closes at device level into an extended diode equation, $J = G_n\delta_0^{(p)} + G_p\delta_0^{(n)}$, which contains the Shockley equation as its low-injection, long-base limit and reproduces the $\mathrm{n} = 1 \rightarrow 2$ high-injection roll-off intrinsically. The junction voltage emerges as a self-limiting drop confined to the space-charge region, separated from the applied bias by exact base voltage drops. As device-level consequences, we quantify the redistribution of power dissipation from the junction into the quasi-neutral base (75–90% at high current) and the field-assisted acceleration of minority carriers, which shortens the base transit time by up to a factor of six. Three parameter-free, experimentally testable predictions follow: a universal ideality-factor transit $\mathrm{n} = (2\delta_0 + N)/(\delta_0 + N)$, a doping-set high-injection onset voltage $V_j^* = 2V_T\ln(N/2n_i)$, and a geometry factor $\coth(W/L_a)$ that survives into high injection.

# I. Introduction

The p–n junction remains the basic building block of semiconductor devices, and operating conditions demanding high bias and high-level injection (in power rectifiers, solar cells, and high-current switches) are increasingly common. It is under these conditions that the classical theory loses its quantitative reliability.

The classical theory, established primarily through Shockley's 1949 formulation [1,2], has been the standard framework of semiconductor device physics for more than seven decades [3-7]. It derives the ideal diode equation and reproduces the terminal current–voltage characteristics accurately under low injection [1,3]. The theory rests, however, on an identifiable simplification of the quasi-neutral base: the majority carriers are assumed to stay at their equilibrium concentration, the base electric field is neglected, and the terminal current is obtained from minority-carrier diffusion evaluated at the space-charge region (SCR) boundary of each base [1,3,4]. Three consequences follow: the theory leaves unexplained how the base carries any net current when the majority carriers are inert; it provides no internal field profile; and it needs an ad hoc external series resistance to reproduce the high-injection roll-off [3,8,9]. That roll-off, the drift of the ideality factor from $n = 1$ toward $n = 2$, has been documented for decades in power rectifiers and p–i–n structures [10–12], and case-specific analyses exist for particular high-current regimes [11-14]. A unified junction law is still needed: one that identifies which assumption fails, repairs it from first principles, contains the Shockley equation as a demonstrable limit, and makes quantitative, falsifiable predictions beyond it. The need is also practical, since the power consumption and the switching speed of conventional junction devices are designed and optimized on that basis.

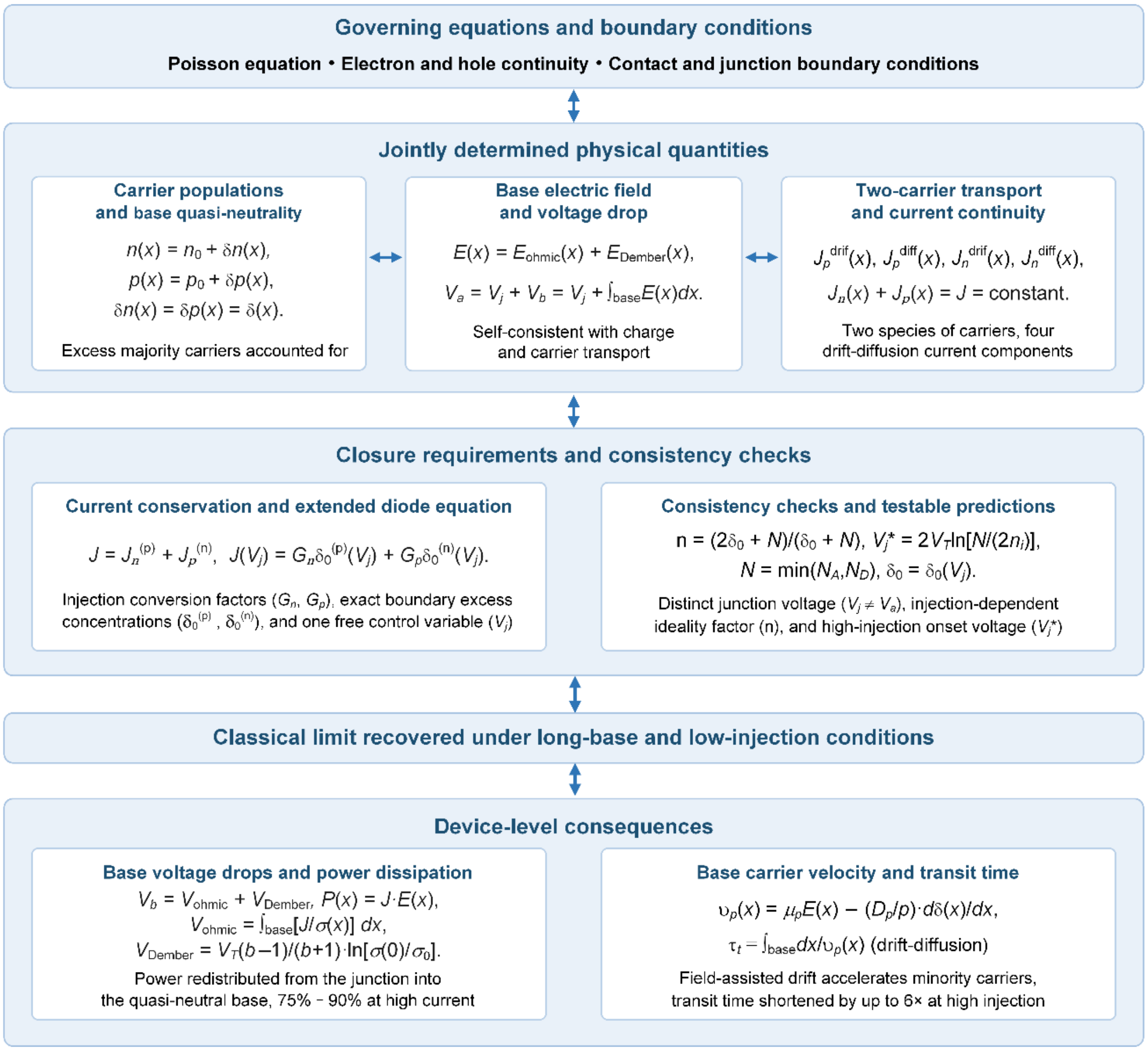

**FIG. 1.** Logical structure of the self-consistent two-carrier framework. The excess electrons and holes share one profile $\delta(x)$ by quasi-neutrality; together with the base field $E(x)$ and the net current $J$ they are locked by the two continuity equations and the two boundary conditions, leaving the junction voltage $V_j$ as the single free parameter that closes the device solution: the profile $\delta(x)$, the field $E(x, V_j)$, the four current components summing to a position-independent $J$, the voltage partition $V_a = V_j + V_b$, and the extended diode equation $J(V_j)$ (Secs. II–VI). The device-level consequences, power dissipation (Sec. VII) and base transit time (Sec. VIII), follow from the field $E(x)$ and the partition. The bottom box recalls, for contrast, the premises of the classical theory.

Our previous work [15] drew attention to the missing piece: an injected minority-carrier population cannot coexist with a quasi-neutral base unless an equal excess majority-carrier population accompanies it, and that excess majority population is a genuine transport participant. Unlike the classical treatment, which addresses only the injection and diffusion of minority carriers, the present framework accounts for the injection, diffusion, and drift of both carrier species, without introducing any assumption beyond the classical ones. Starting from the steady-state continuity equations, the drift–diffusion equations, and Poisson's equation, we retain minority-carrier drift throughout (a term that earlier treatments could not close), derive a boundary relation that couples the two SCR boundaries through a single junction voltage $V_j$, and obtain closed-form expressions for all four current components as functions of position. Their sum is a position-independent current, and the resulting diode equation holds uniformly across injection level, bias polarity, and base length. Applied to the two device-level quantities that the classical theory can only treat with empirical patches, the framework reverses both classical errors: at fixed bias the classical theory overestimates the dissipation but underestimates it at fixed current, and at high injection it overestimates the base transit time by up to a factor of six.

Throughout, the junctions are uniformly doped abrupt structures and the notation follows that of mainstream textbooks [3,4,6,16,17]. All derivations are performed for the n-type base unless otherwise indicated; the p base is fully symmetric. Low- and high-injection conditions are treated on the same footing, and the classical limit is recovered explicitly wherever the framework is applied.

The paper is organized to mirror the physical logic of the framework (Fig. 1): Secs. II–VI develop the two-carrier description and close it into the extended diode equation; Secs. VII and VIII treat power dissipation and base transit time; Sec. IX discusses the relation to earlier high-injection treatments, the validity boundaries, the implications for device modeling, and an experimental verification route based on existing diode data; Sec. X concludes. A condensed account of the framework and of the extended diode equation is presented in a companion Letter [18].

# II. Excess majority carriers and base quasi-neutrality

## A. The excess majority carriers

The starting point of the framework is the relation

$$\delta n(x) = \delta p(x) \equiv \delta(x), \tag{1}$$

where $\delta(x)$ denotes the excess carrier concentration at position $x$ in the base. Taking the n-type base for analysis, it was proven in our previous work [15], by both numerical simulation of the full coupled drift–diffusion–Poisson system and analytical derivation, that there exist excess majority carriers (EMaCs, $\delta n(x)$) and an equal population of excess minority carriers (EMiCs, $\delta p(x)$) at the same position in the base: the relative imbalance $(\delta p - \delta n)/\delta p$ is of the order of $\tau_d/\tau_p$, where $\tau_d = \varepsilon/\sigma$ is the dielectric relaxation time, $\tau_p$ the minority-carrier (hole) lifetime, and $\sigma$ the conductivity. For the materials commonly used in p–n junctions, the ratio $\tau_d/\tau_p$ (with doping of $10^{16}$ cm$^{-3}$) is $\sim 5 \times 10^{-8}$ (Si), $\sim 2 \times 10^{-8}$ (Ge), $\sim 8 \times 10^{-5}$ (GaAs), $\sim 6 \times 10^{-7}$ (4H-SiC), $\sim 1 \times 10^{-4}$ (GaN) [3,19,20], typically $10^{-7}$–$10^{-5}$, therefore it is far below any transport-relevant magnitude, so that $\delta p(x) - \delta n(x) \approx 0$. The simulations of Ref. [15] confirm that the relation holds for both forward and reverse bias and for both low and high injection. These are first-principles bounds for the existence of excess majority carriers in view of dielectric relaxation, and they justify imposing $\delta p(x) = \delta n(x)$, not as an approximation, at the level of the analytical framework.

## B. The base quasi-neutrality

Charge quasi-neutrality requires the net space charge at position $x$ to be nearly zero. In the presence of excess carriers,

$$\rho(x) = q[N_d(x) - N_a(x) + p_0(x) - n_0(x) + \delta p(x) - \delta n(x)] = q[\delta p(x) - \delta n(x)] \approx 0,$$

$$\delta n(x) \approx \delta p(x), \tag{2}$$

in agreement with the first-principles bound above. That bound also reveals the physical origin of base quasi-neutrality. Classical theory invokes the equality $\delta n = \delta p$ only under high-level injection [3,4]; we assert, and the dielectric-relaxation argument confirms, that it holds under arbitrary operating conditions. Poisson's equation ties any nonzero $\rho$ to a diverging field: if $\delta n$ were absent while $\delta p$ were injected into the n base, $\rho = q\delta p$ would be of the order of the doping density, severely violating the observed quasi-neutrality. The only way to inject holes into the n base without catastrophic space charge is for an equal electron population to appear alongside them. The existence of EMaCs is therefore not optional; it is *forced* by the same electrostatics that defines a quasi-neutral region.

The residual $(\delta p - \delta n)$ is not the driver of the base field; it is its *electrostatic shadow*. The field is fixed by the transport equations; Poisson's equation then merely reports the minute charge separation that accompanies it. Writing the causality the other way, with a tiny charge difference generating the field [4], is physically backwards. We therefore impose quasi-neutrality as $\delta p(x) = \delta n(x)$ and let the field follow from transport.

# III. The excess-carrier profile

## A. The ambipolar diffusion equation

The excess-carrier profile $\delta(x)$ in the base follows from the full drift–diffusion system. Quasi-neutrality, $\delta n(x) = \delta p(x) \equiv \delta(x)$, implies that the two species share identical spatial gradients, $dn/dx = dp/dx = d\delta/dx$, for the uniformly doped base. The electron and hole current densities are

$$J_n = q\mu_n nE + qD_n \frac{d\delta}{dx}, \qquad J_p = q\mu_p pE - qD_p \frac{d\delta}{dx}, \tag{3}$$

with $n = n_{n0} + \delta$ and $p = p_{n0} + \delta$. The steady-state continuity equations, with Shockley–Read–Hall recombination [21,22] as the only bulk process in the base, are

$$\frac{1}{q}\frac{dJ_n}{dx} = \frac{\delta}{\tau_n}, \qquad -\frac{1}{q}\frac{dJ_p}{dx} = \frac{\delta}{\tau_p}. \tag{4}$$

Inserting the current expressions, multiplying the first equation by $\mu_p p$ and the second by $\mu_n n$, and adding, the terms containing $dE/dx$ combine as $\mu_n\mu_p(np - np)\, dE/dx = 0$ and cancel exactly, leaving the ambipolar transport equation [23-26] with the drift term still present:

$$D_a \frac{d^2\delta}{dx^2} + \mu_a E \frac{d\delta}{dx} = \frac{\delta}{\tau_a}, \tag{5}$$

$$D_a = \frac{D_p\mu_n n + D_n\mu_p p}{\mu_n n + \mu_p p}, \qquad \mu_a = \frac{\mu_n\mu_p(p - n)}{\mu_n n + \mu_p p}, \qquad \tau_a = \frac{\mu_n n + \mu_p p}{\mu_n n/\tau_p + \mu_p p/\tau_n},$$

where $D_a$, $\mu_a$, and $\tau_a$ are the ambipolar diffusivity, mobility, and lifetime; all three reducing to the minority-carrier values $D_p$, 0, and $\tau_p$ at low injection, while at high injection $D_a \to 2D_nD_p/(D_n+D_p)$, $\mu_a \to 0$, and $\tau_a \to \tau_n\tau_p(\mu_n+\mu_p)/(\mu_p\tau_p+\mu_n\tau_n)$ (identically $\tau$ for the equal-lifetime model used throughout).

Two features of the ambipolar coefficients are worth noting. First, $D_a$ is not the minority-carrier diffusivity $D_p$: the internal field that maintains quasi-neutrality speeds up the slower species and retards the faster one, and this mutual electrostatic coupling is already folded into $D_a$; this is the sense in which "using ambipolar diffusion" *does* incorporate the field's effect on the transport of the neutral excess-carrier package. Second, the term $\mu_a E\, d\delta/dx$, the net drift of that package, is a genuinely distinct effect governed by the separate coefficient $\mu_a$.

This net-drift term is second-order small and can be treated as a perturbation [24,25,27], yielding the ambipolar diffusion equation

$$\frac{d^2\delta}{dx^2} = \frac{\delta}{L_a^2}, \qquad L_a = \sqrt{D_a \tau_a}, \tag{6}$$

with $L_a$ the ambipolar diffusion length. The elimination is used *only* to determine the profile $\delta(x)$; it does not mean that the electric field is assumed zero. The framework remains internally consistent: the ambipolar reduction absorbs the field's coupling into $D_a$, while the physically real drift currents of both species are retained in full when the individual components are computed (Sec. V). This is the point at which the present framework departs from the classical treatment, which uses $D_p$ for the profile *and* discards the minority-carrier drift current altogether. A quantitative bound on the neglected term is derived in Appendix A: the associated ambipolar Péclet number is capped at $1/4(b+1) \approx 6\%$ for Si over the entire injection range, vanishing in both the low- and high-injection limits.

## B. The hyperbolic profile

We denote the n-side SCR boundary by $x = 0$ and the ohmic contact by $x = W$, with boundary conditions

$$\delta(0) = \delta_0(V_j), \qquad \delta(W) = 0.$$

The solution is a unified hyperbolic profile as [3,23,24]

$$\delta(x) = \delta_0 \frac{\sinh\left[\frac{W-x}{L_a}\right]}{\sinh\left(\frac{W}{L_a}\right)}, \tag{7}$$

whose gradient is likewise hyperbolic,

$$\frac{d\delta(x)}{dx} = -\frac{\delta_0}{L_a} \frac{\cosh\left[\frac{W-x}{L_a}\right]}{\sinh\left(\frac{W}{L_a}\right)}. \tag{8}$$

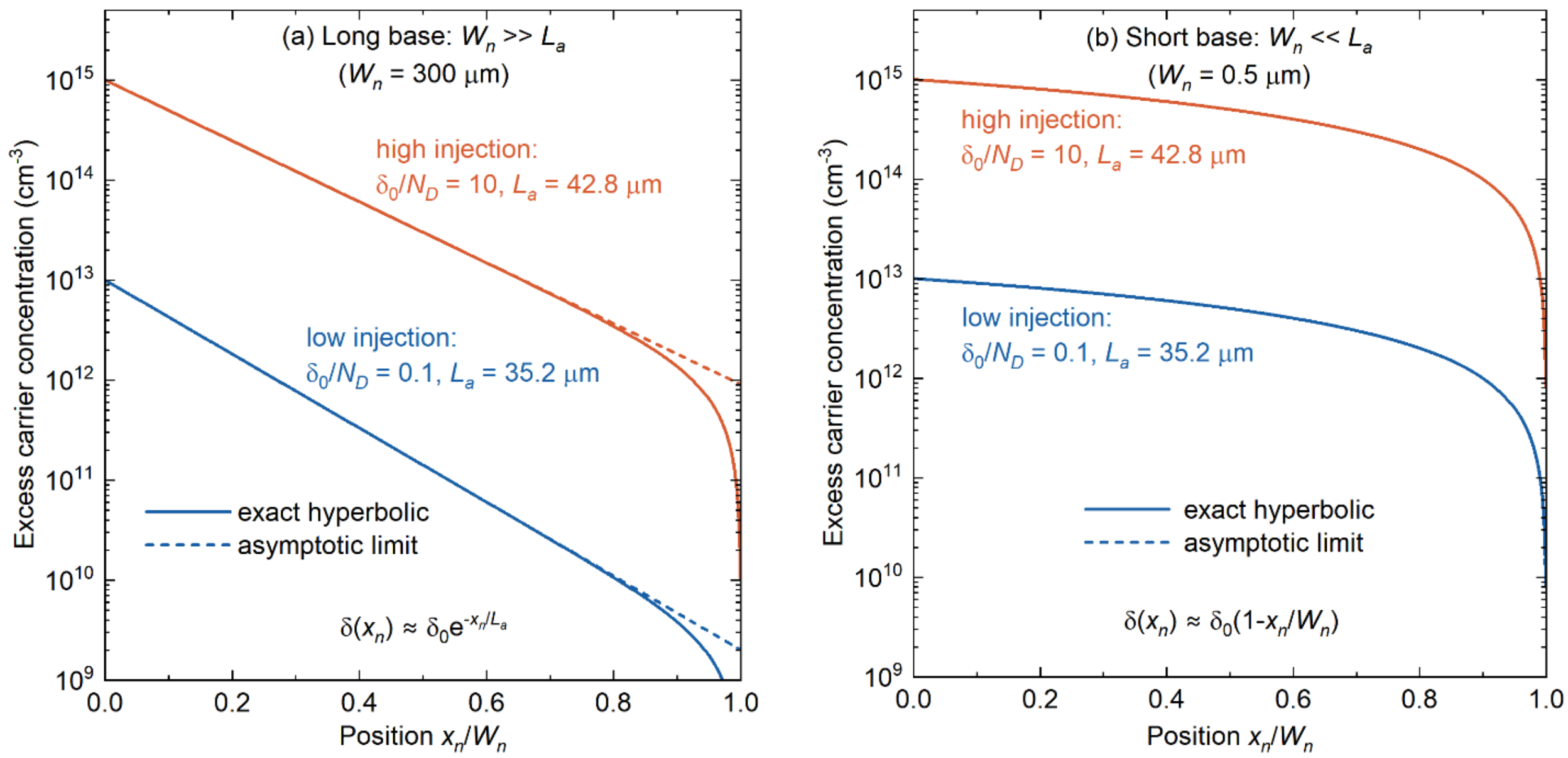

**FIG. 2.** Excess-carrier profiles $\delta(x)$ in the n-type base from the closed-form solution of the ambipolar diffusion equation, with $\delta(0) = \delta_0$ and $\delta(W) = 0$. Solid curves are the exact hyperbolic profiles; dashed curves are the asymptotic limits. Two injection levels are shown, set by the boundary value $\delta_0$ (Sec. III C): low injection $\delta_0/N_D = 0.1$ and high injection $\delta_0/N_D = 10$ ($N_D = 10^{14}$ cm$^{-3}$, $\tau = 1\ \mu$s), with the appropriate diffusion length in each case ($L_a \rightarrow L_p = 35.2\ \mu$m at low injection, $L_a = 42.8\ \mu$m at high injection). (a) Long base ($W = 300$ $\mu$m $\gg L_a$): the profile reduces to the exponential decay over essentially the whole base, deviating only in the immediate vicinity of the contact. (b) Short base ($W = 0.5\ \mu$m $\ll L_a$): recombination is negligible and the profile becomes nearly linear at both injection levels.

As Fig. 2 shows, the single profile covers both the long-base ($W \gg L_a$) and short-base ($W \ll L_a$) regimes without separate case analysis: the former reduces to the textbook exponential $\delta \rightarrow \delta_0 e^{-x/L_a}$ with $\coth(W/L_a) \rightarrow 1$ [1,3], the latter to the linear form $\delta \rightarrow \delta_0(1 - x/W)$ with $\coth(W/L_a) \rightarrow L_a/W$ [3,4].

A remark on the role of the diffusivities is in order. Each carrier's diffusion current must use its own diffusion coefficient $D_n$ or $D_p$, since $D$ reflects the intrinsic transport mobility of that carrier; the ambipolar $D_a$ is an effective parameter of the *coupled* distribution and cannot represent individual carrier fluxes. The physical essence is that collective behavior (via $D_a$ and $\tau_a$) determines the distribution shape, parametrized by $L_a$, while individual carrier currents through that distribution are determined by each carrier's intrinsic transport properties (Sec. V).

## C. The exact boundary excesses

The boundary value $\delta_0$ at either SCR boundary is not an independent parameter; it is fixed by the junction law through the single junction voltage $V_j$. Applying the law of mass action and quasi-neutrality to both carrier species at both SCR boundaries gives the exact boundary relations [15]

$$\delta_0^{(n)}(V_j) = \left[\left(\frac{N_D}{2}\right)^2 + n_i^2\, e^{V_j/V_T}\right]^{\frac{1}{2}} - \frac{N_D}{2} - p_{n0}, \tag{9-1}$$

$$\delta_0^{(p)}(V_j) = \left[\left(\frac{N_A}{2}\right)^2 + n_i^2\, e^{V_j/V_T}\right]^{\frac{1}{2}} - \frac{N_A}{2} - n_{p0}, \tag{9-2}$$

which are accurate for arbitrary bias and injection level [15], as shown in Fig. 3. Combining and subtracting the two expressions yields the coupled relation

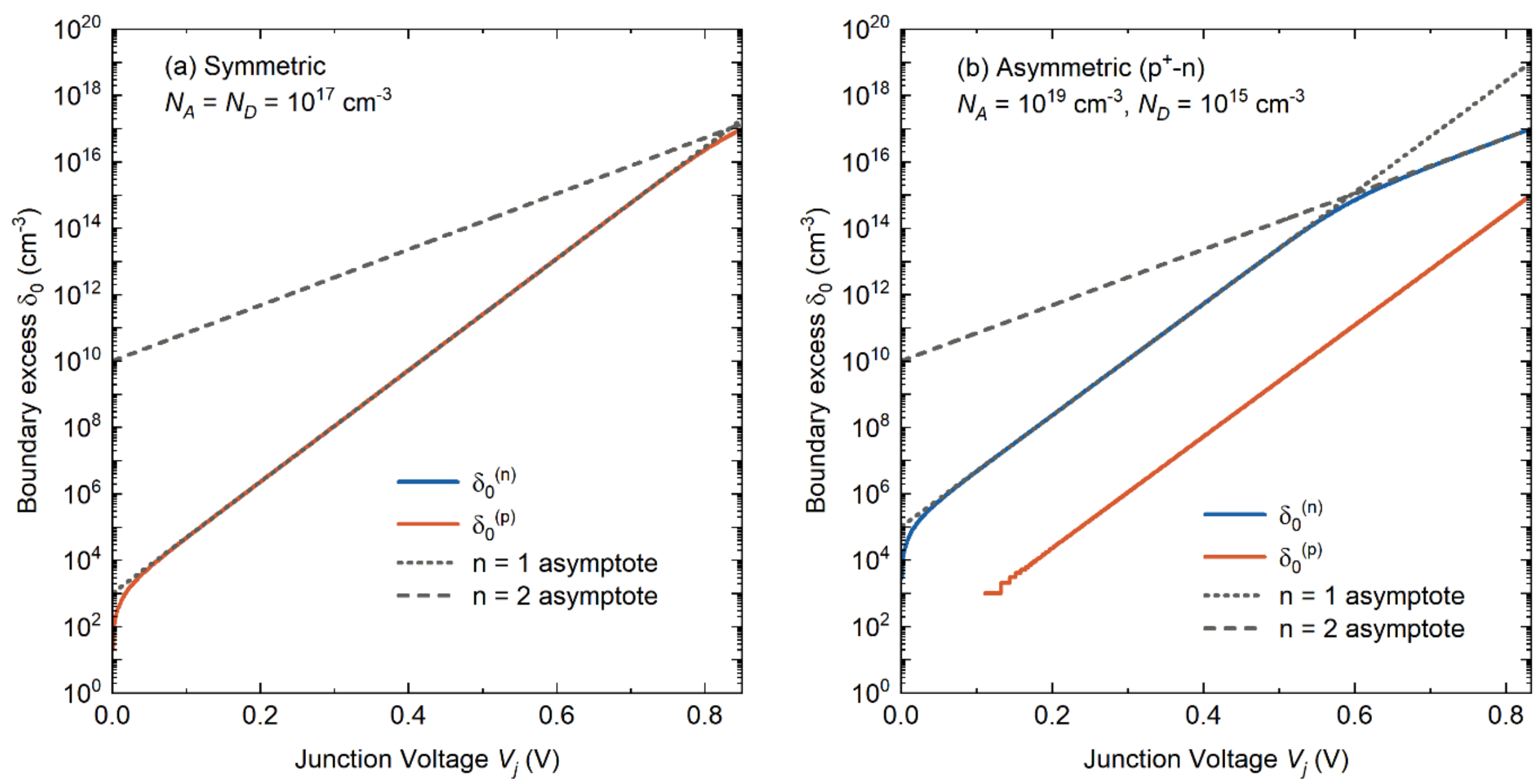

**FIG. 3.** Boundary excess concentrations at the two SCR edges, $\delta_0^{(n)}$ and $\delta_0^{(p)}$, as functions of the common junction voltage $V_j$, the origin of the boundary-to-boundary locking of the two carrier populations. (a) Symmetric junction ($N_A = N_D = 10^{17}$ cm$^{-3}$): the two excess concentrations coincide at every bias, following the low-injection (n = 1, dashed) asymptote at small $V_j$ and the high-injection (n = 2, dash-dotted) asymptote $\delta_0 \approx n_i e^{V_j/2V_T}$ at large $V_j$. (b) Asymmetric p$^+$–n junction ($N_A = 10^{19}$ cm$^{-3}$, $N_D = 10^{15}$ cm$^{-3}$): at low injection the two boundaries are strongly unequal, each following its own doping-dependent n = 1 asymptote; as $V_j$ increases, the lightly doped side approaches the n = 2 asymptote while the other does not. Because $N_A N_D = 10^{34}$ cm$^{-6}$ for both configurations, the built-in voltage $V_0 = V_T \ln(N_A N_D / n_i^2) = 0.833$ V (dotted vertical line) is shared, marking the upper limit of $V_j$.

$$\left(\delta_0^{(n)} + \frac{N_D}{2} + p_{n0}\right)^2 - \left(\delta_0^{(p)} + \frac{N_A}{2} + n_{p0}\right)^2 = \left(\frac{N_D}{2}\right)^2 - \left(\frac{N_A}{2}\right)^2, \quad (10)$$

which expresses the boundary-to-boundary locking of the two excess populations: both are controlled simultaneously by the single junction voltage and cannot be set independently.

# IV. The base electric field and structural voltage drops

## A. The self-consistent base field

The electric field in the quasi-neutral base is determined self-consistently from the current definitions. Adding the electron and hole currents,

$$J = J_n + J_p = q(\mu_n n + \mu_p p)E + q(D_n - D_p)\frac{d\delta}{dx}, \quad (11)$$

and solving for the field,

$$E(x) = \frac{J - q(D_n - D_p)\frac{d\delta}{dx}}{q(\mu_n n + \mu_p p)} = \frac{J}{\sigma(x)} - \frac{q(D_n - D_p)}{\sigma(x)}\frac{d\delta(x)}{dx} \equiv E_{\text{ohmic}} + E_{\text{Dember}}, \quad (12)$$

where the position-dependent conductivity is

$$\sigma(x) = q(\mu_n n_{n0} + \mu_p p_{n0}) + q(\mu_n + \mu_p)\delta(x) = \sigma_0 + q(\mu_n + \mu_p)\delta(x),$$

with $\sigma_0 = q(\mu_n n_{n0} + \mu_p p_{n0})$ its equilibrium value. The quantity $J$ is the net current, by conservation a position-independent constant rather than an externally imposed variable. The base field thus separates into an ohmic drift term, associated with the conductivity modulation of the base, and a Dember counter-field, driven by the ambipolar gradient and the unequal electron and hole diffusivities [28,54-56]. Substituting the hyperbolic profile gives the explicit form

$$E(x) = \frac{J + q(D_n - D_p)\frac{\delta_0}{L_a}\frac{\cosh[(W-x)/L_a]}{\sinh(W/L_a)}}{\sigma_0 + q(\mu_n + \mu_p)\,\delta_0\frac{\sinh[(W-x)/L_a]}{\sinh(W/L_a)}}. \quad (13)$$

As shown in Fig.4, in all operating regimes $E(x)$ is nonzero throughout the quasi-neutral base, in direct contrast to the classical assumption $E = 0$; a fully closed form is obtained once the net current $J$ is determined (Sec. VI).

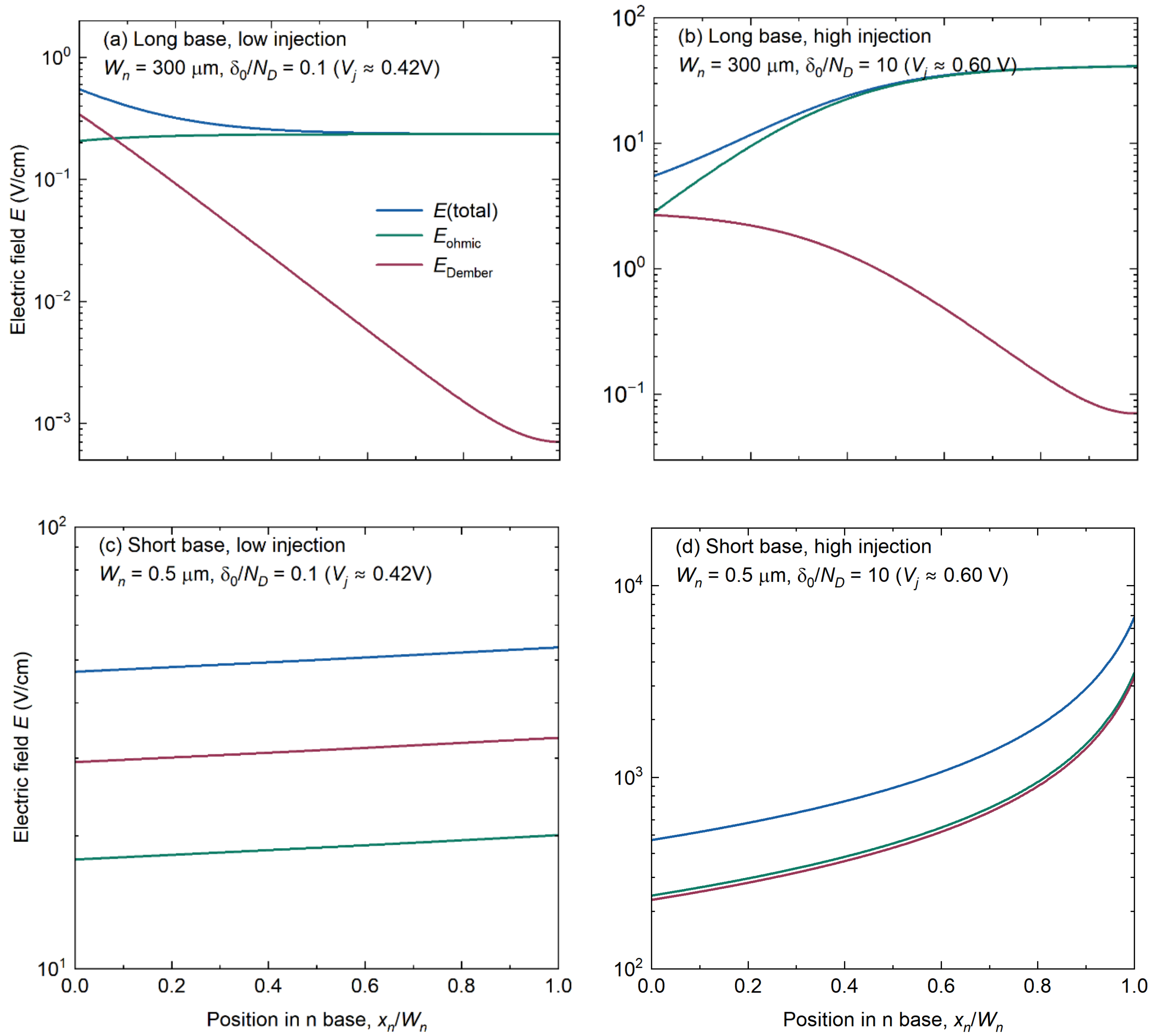

**FIG. 4.** Base electric field $E(x)$ in the n-type base and its decomposition into the ohmic part $E_{\text{ohmic}} = J/\sigma(x)$, associated with the conductivity modulation $\sigma(x) = \sigma_0 + q(\mu_n + \mu_p)\delta(x)$, and the Dember counter-field $E_{\text{Dember}} = -q(D_n - D_p)\delta'(x)/\sigma(x)$. The total current $J$ is fixed by the closure relation of the asymmetric p$^+$–n junction ($N_D = 10^{14}$ cm$^{-3}$, $L_a = 42.8\ \mu$m). (a) Long base ($W = 300\ \mu$m), low injection ($\delta_0/N_D = 0.1$, $V_j \approx 0.42$ V): the ohmic term dominates over most of the base; the Dember field is appreciable only near the junction. (b) Long base, high injection ($\delta_0/N_D = 10$, $V_j \approx 0.60$ V): the Dember field becomes comparable to the ohmic field near the junction, while the ohmic field rises steeply toward the unmodulated end of the base. (c) Short base ($W = 0.5\ \mu$m), low injection: both components are spatially uniform. (d) Short base, high injection: the two components are comparable over the whole base and grow toward the end contact.

## B. The junction voltage

The existence of a base field necessarily implies a voltage drop across the base; however, the classical theory cannot solve for it. Neglecting the ohmic drops at the end contacts, the applied bias $V_a$ distributes itself over three series regions: the SCR, carrying the junction voltage drop $V_j$, and the two bases, carrying the base voltage drops $V_b^{(n)}$ and $V_b^{(p)}$:

$$V_a = V_j + V_b^{(n)} + V_b^{(p)}. \tag{14}$$

The classical theory conflates $V_j$ with $V_a$ [1,3] as a consequence of the assumption $E = 0$. In fact, $V_j$ is solely due to the presence of non-equilibrium excess carriers: it is an internal quantity supported by the population and distribution of excess carriers through an electrostatic mechanism. This excess-induced voltage superposes on the built-in potential of the SCR, changing the electrostatic potential difference between the two boundaries from its equilibrium value $V_0$ to the non-equilibrium value $V_0 - V_j$; the change is what we call the junction voltage. $V_j$ is thereby cleanly separated from $V_a$ and bounded by $V_0$, to which it is confined as a self-limiting drop.

This separation has a broader significance. Taking equilibrium as the reference, whenever excess carriers are present (whatever their origin), a corresponding junction voltage arises: the junction voltage is a consequence of the excess carriers, and the applied bias is only one of several possible causes of the latter. A junction voltage can exist with no applied bias at all, as in the photovoltaic effect under illumination [3,29]; and it persists transiently after the bias is withdrawn, decaying as the stored excess charge recombines, as in the reverse-recovery transient of power diodes [8,9,30].

## C. The base voltage drops

The identification of $E(x)$ enables the base voltage drop to be evaluated exactly. Integrating the field across the base,

$$V_b = \int_0^W \frac{J}{\sigma(x)}\,dx + V_T\frac{b-1}{b+1}\ln\frac{\sigma(0)}{\sigma_0} \equiv V_{\text{ohmic}} + V_{\text{Dember}} \qquad \text{(for either base)}, \tag{15}$$

where $b = \mu_n/\mu_p = D_n/D_p$ (Einstein relation applied). $V_b$ separates into an ohmic drop and a profile-independent Dember drop. The applied bias therefore partitions exactly as

$$V_a = V_j + V_{\text{ohmic}}^{(n)} + V_{\text{Dember}}^{(n)} + V_{\text{ohmic}}^{(p)} + V_{\text{Dember}}^{(p)}. \tag{16}$$

The partition is illustrated in Fig. 5: $V_j$ falls progressively below $V_a$ as injection rises, because part of the bias is consumed in driving the quasi-neutral base field and sustaining the majority-carrier current, and it saturates smoothly toward $V_0$, which it cannot exceed. The separation is exact and requires no limiting approximation; its quantitative consequences for the device power budget are developed in Sec. VII.

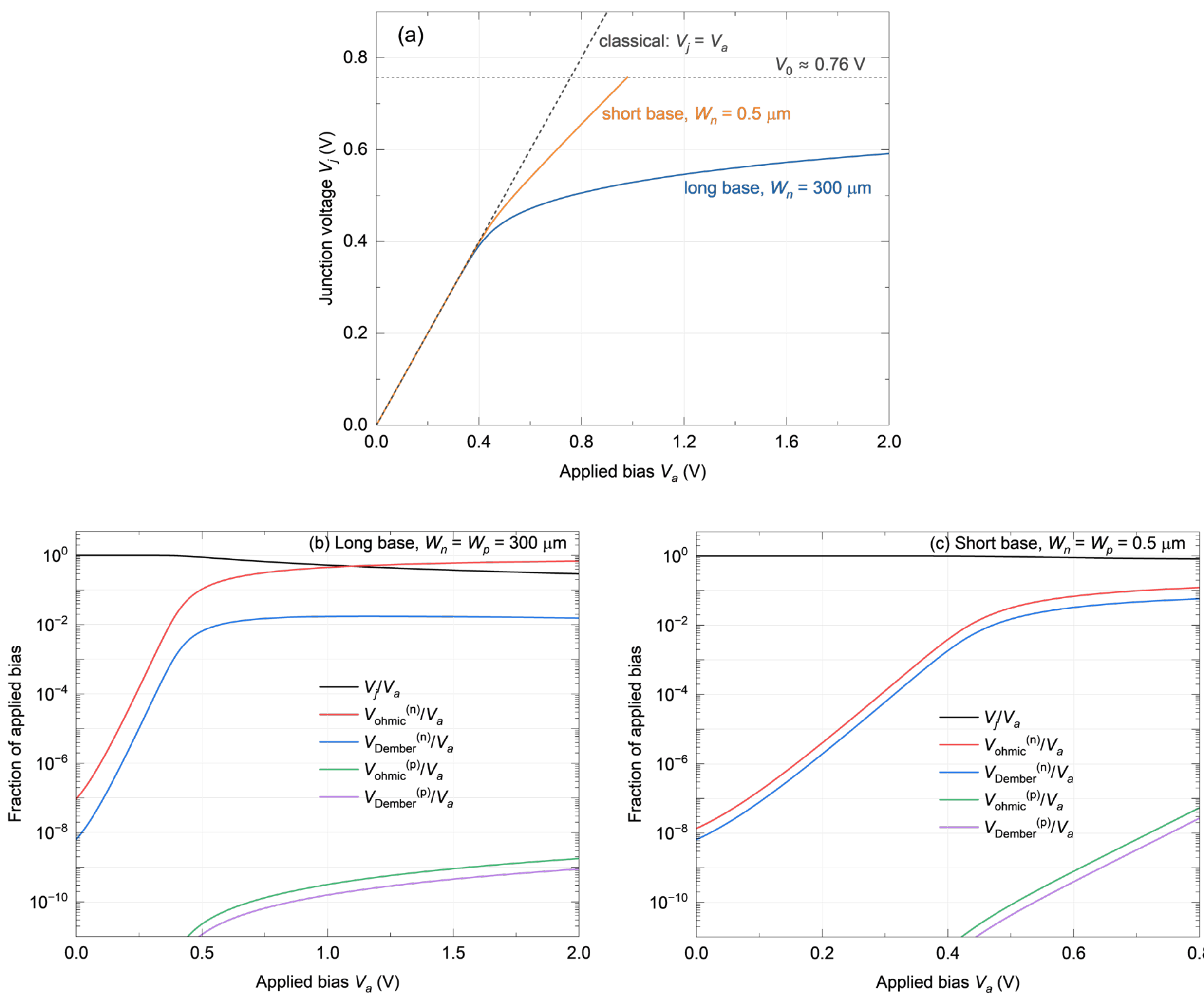

**FIG. 5.** Partition of the applied bias over the three series regions, $V_a = V_j + V_b^{(n)} + V_b^{(p)}$, each base drop separating exactly into ohmic and Dember components. (a) Junction voltage $V_j$ versus applied bias $V_a$ for the long-base ($W = 300\ \mu$m) and short-base ($W = 0.5\ \mu$m) devices. The classical treatment conflates $V_j$ with $V_a$ (dashed line); here $V_j$ falls progressively below $V_a$ as injection rises and saturates toward the built-in potential $V_0 = 0.756$ V (dotted line). (b), (c) Fractions of the applied bias carried by the five partition terms for the (b) long-base and (c) short-base devices; the fractions sum to unity at every bias. In the long-base device the n-base drops become appreciable already at moderate injection and dominate at high injection (strong conductivity modulation of the lightly doped, wide base), whereas in the short-base device $V_j \approx V_a$ holds to much higher bias, quantifying when and why the classical identification $V_j = V_a$ breaks down. Parameters: asymmetric $p^+$–n junction, $N_A = 5 \times 10^{18}$ cm$^{-3}$ (p base, $\mu_n = 150$, $\mu_p = 50$ cm$^2$/V·s, $L_a = 13.9\ \mu$m, $W_p = 1\ \mu$m), $N_D = 10^{14}$ cm$^{-3}$ (n base, $\mu_n = 1350$, $\mu_p = 480$ cm$^2$/V·s, $L_a = 42.8\ \mu$m).

# V. Current components and current continuity

## A. The four spatial components

With the profile and its gradient in hand, we denote

$$s(x) = \frac{\sinh[(W-x)/L_a]}{\sinh(W/L_a)},$$

$$c(x) = \frac{\cosh[(W-x)/L_a]}{\sinh(W/L_a)},$$

$$\sigma(x) = \sigma_0 + q(\mu_n + \mu_p)\delta_0\, s(x).$$

The field and the four current components follow in closed form:

$$E(x) = \frac{J + q\left(D_n - D_p\right)\frac{\delta_0}{L_a}\, c(x)}{\sigma_0 + q\left(\mu_n + \mu_p\right)\delta_0\, s(x)}, \tag{17}$$

$$J_n^{\mathrm{diff}}(x) = qD_n\frac{d\delta}{dx} = -qD_n\frac{\delta_0}{L_a}c(x), \tag{18 − 1}$$

$$J_p^{\mathrm{diff}}(x) = -qD_p\frac{d\delta}{dx} = qD_p\frac{\delta_0}{L_a}c(x), \tag{18 − 2}$$

$$J_n^{\mathrm{drift}}(x) = q\mu_n[n_{n0} + \delta_0 s(x)]E(x), \tag{18 − 3}$$

$$J_p^{\mathrm{drift}}(x) = q\mu_p[p_{n0} + \delta_0 s(x)]E(x). \tag{18 − 4}$$

Each species uses its own diffusivity for its individual contribution, in accordance with the remark of Sec. III B. Three of the four components are missing from the classical theory; here both carrier species participate in transport by diffusion and drift, weighted by their local populations.

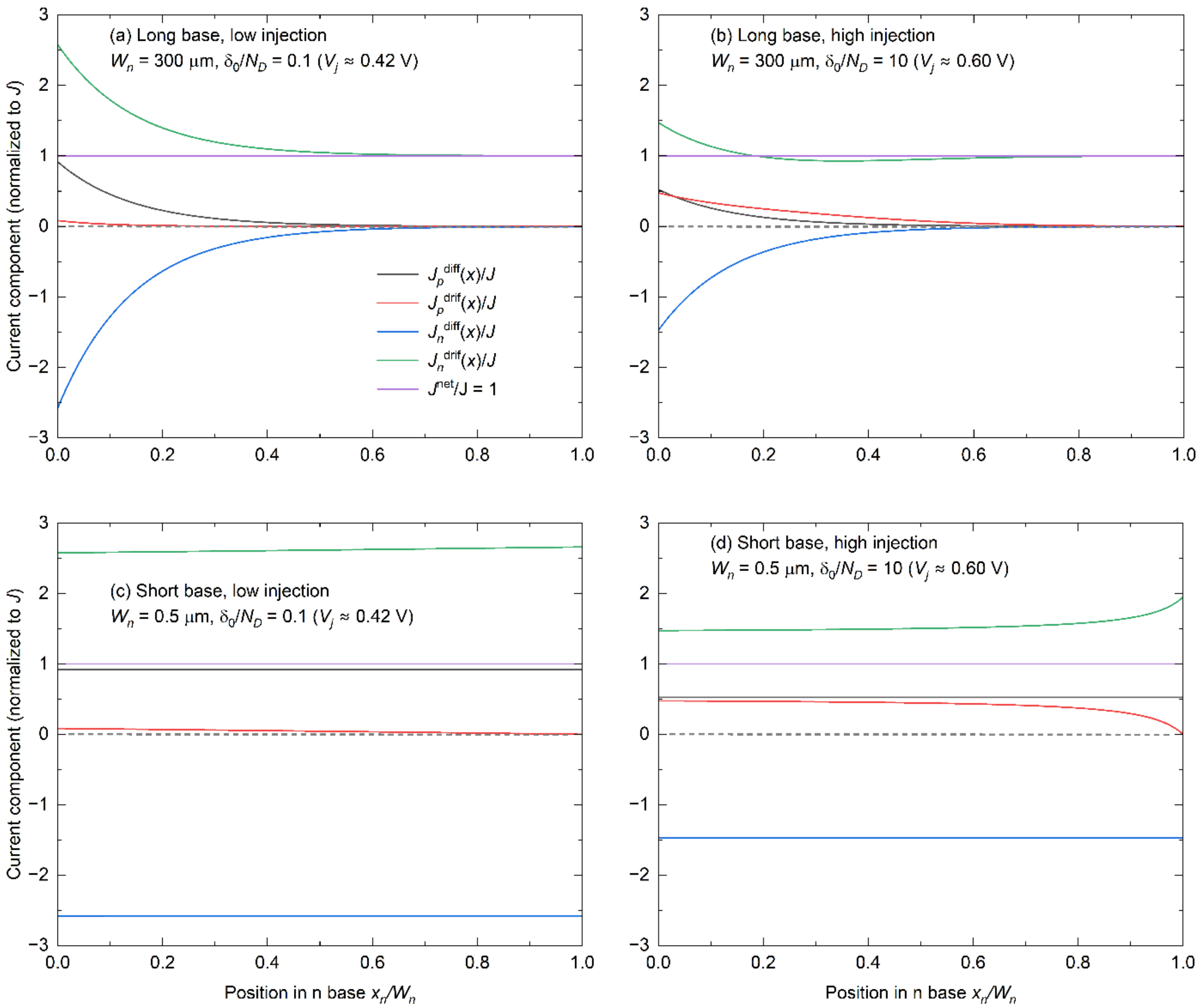

**FIG. 6.** Spatial current components in the n-base and their continuity, all normalized to the total current $J$. The position-dependent parts cancel exactly, so $J(x) = J$ at every plane. (a) Long base ($W = 300$ μm), low injection ($\delta_0/N_\mathrm{D} = 0.1$, $V_j \approx 0.42$ V, $J = 5.07 \times 10^{-3}$ A/cm$^2$). At the SCR edge the current is carried almost entirely by the injected holes, $J_p^{\mathrm{diff}}(0) = 0.92J$ and $J_p^{\mathrm{drift}}(0) = 0.08J$, while the electron back-diffusion $J_n^{\mathrm{diff}}(0) = -2.58J$ is exactly offset by electron drift $J_n^{\mathrm{drift}}(0) = +2.58J$. Both diffusion components decay on the scale of $L_\mathrm{a}$, and the current is progressively transferred to majority-electron drift, which carries essentially the whole current at the end contact: a continuous hole-to-electron current conversion at constant total current. (b) Long base, high injection ($\delta_0/N_\mathrm{D} = 10$, $V_j \approx 0.60$ V, $J = 0.887$ A/cm$^2$): conductivity modulation makes hole drift no longer negligible (52% diffusion, 48% drift at $x = 0$), and the normalized electron counter-flow shrinks to $\pm 1.47J$. (c) Short base ($W = 0.5$ μm), low injection ($\delta_0/N_\mathrm{D} = 0.1$, $V_j \approx 0.42$ V): with $W \ll L_\mathrm{a}$ every component is nearly position independent, the electron back-flow ($\approx \pm 2.5J$) locked flat across the base and the hole current ($\approx 0.97J$) carried almost entirely by diffusion. (d) Short base, high injection ($\delta_0/N_\mathrm{D} = 10$, $V_j \approx 0.60$ V): hole drift is again comparable to hole diffusion near the junction (each $\approx 0.45J$) and vanishes at the end contact, the electron counter-flow staying at $\approx \pm 1.5J$. In all four cases, the majority-electron drift is the largest component throughout the base.

## B. The current continuity

The net current at position $x$ is the sum of the four components at the same position. The drift pair sums to

$$J_n^{\mathrm{drift}}(x) + J_p^{\mathrm{drift}}(x) = \sigma(x)E(x) = J + q\left(D_n - D_p\right)\frac{\delta_0}{L_a}c(x), \tag{19}$$

and the diffusion pair sums to exactly its negative,

$$J_n^{\mathrm{diff}}(x) + J_p^{\mathrm{diff}}(x) = -q\left(D_n - D_p\right)\frac{\delta_0}{L_a}c(x). \tag{20}$$

The position-dependent parts therefore cancel identically:

$$J^{\mathrm{net}}(x) = \left[J + q\left(D_n - D_p\right)\frac{\delta_0}{L_a}c(x)\right] - \left[q\left(D_n - D_p\right)\frac{\delta_0}{L_a}c(x)\right] = J = \mathrm{const.} \tag{21}$$

As shown in Fig.6, although each component is position dependent, the net current is a position-independent constant: current continuity is exhibited explicitly at every plane, not asserted from global charge conservation, and not approximated at special points. Near the SCR edge the current is carried almost entirely by the injected minority carriers (92% by hole diffusion at low injection); moving into the base it is progressively transferred to majority-carrier drift, the large electron back-diffusion near the junction ($-2.58J$) being exactly offset by electron drift ($+2.58J$), so that the net electron current vanishes at the SCR edge of the asymmetric junction, as it must. The classical theory merely asserts continuity; here it is demonstrated through the four components, three of which are absent from the classical treatment.

# VI. The extended diode equation

## A. Current closure and the injection conversion factors

The net current $J$ that appears in the base field and in the current components is not an externally imposed variable; it is fixed by current conservation at the two SCR boundaries. Generation–recombination in the SCR is neglected, as in the classical theory [1], so the injected electron and hole currents are each conserved across the SCR, and the net current equals the sum of the two minority-carrier injection currents:

$$J = J_n\left(x_p = 0\right) + J_p(x_n = 0), \tag{22}$$

where $J_n(x_p = 0)$ is the electron current injected into the p-base and $J_p(x_n = 0)$ the hole current injected into the n-base, each evaluated at its own SCR boundary:

$$J_n(x_p = 0) = \frac{qD_n^{(p)}\delta_0^{(p)}}{L_a^{(p)}}\coth\frac{W_p}{L_a^{(p)}}, \qquad J_p(x_n = 0) = \frac{qD_p^{(n)}\delta_0^{(n)}}{L_a^{(n)}}\coth\frac{W_n}{L_a^{(n)}}, \tag{23}$$

the superscripts labeling the base regions involved. Defining the two *injection conversion factors* (ICFs)

$$G_n \equiv \frac{qD_n^{(p)}}{L_a^{(p)}}\coth\frac{W_p}{L_a^{(p)}}, \qquad G_p \equiv \frac{qD_p^{(n)}}{L_a^{(n)}}\coth\frac{W_n}{L_a^{(n)}}, \tag{24}$$

each of which converts a unit boundary excess concentration into the corresponding current, the net current takes the closed form

$$J(V_j) = G_n\,\delta_0^{(p)}(V_j) + G_p\,\delta_0^{(n)}(V_j) \qquad \text{(extended diode equation).} \tag{25}$$

This is the device-level closure of the framework: the current–voltage characteristic of the p–n junction.

## B. Departures from the classical law

The extended diode equation is assembled through the same closure condition as the classical derivation, that the net current is the sum of the two minority-carrier currents injected across the SCR boundaries, but it departs from the classical law in five respects. (1) *Full drift–diffusion injection currents*: each injection current here is the full minority-carrier current, containing both diffusion and drift, the latter arising from the base field that the classical treatment sets to zero a priori. (2) *A well-defined junction voltage*: the independent variable is $V_j$ rather than $V_a$; the two coincide only in the idealized, injection-independent treatment, and writing the law in terms of $V_j$ bounds the exponential factor $e^{V_j/V_T}$ by the built-in potential, so the current remains finite as injection rises while the base drops are accounted for separately instead of being absorbed into an empirical series resistance. (3) *Exact boundary concentrations*: no low-injection factorization is performed, so the equation remains valid across the entire injection range: from reverse bias, through low injection, to high injection. (4) *Injection conversion factors*: material ($D$), geometry ($W$ through $\coth(W/L_a)$), and field coupling ($L_a$) all enter the current law through $G_n$ and $G_p$. (5)

*Intrinsic high-injection roll-off*: as $V_j$ increases, the boundary excess crosses over from $\propto e^{V_j/V_T}$ to $\propto n_i e^{V_j/2V_T}$, so the ideality factor rises continuously from 1 toward 2 from first principles [10–12], without any fitted series resistance.

## C. Reduction to the classical diode equation

As a first consistency check, the extended equation must reproduce the classical equation where the classical assumptions hold: low injection and long base. Under low injection, $\delta_0 \ll N$, the exact boundary excesses reduce to the Boltzmann form

$$\delta_0^{(p)}(V_j) = n_{p0}\left(e^{V_j/V_T} - 1\right), \qquad \delta_0^{(n)}(V_j) = p_{n0}\left(e^{V_j/V_T} - 1\right), \tag{26}$$

and the extended equation becomes $J = (G_n n_{p0} + G_p p_{n0})(e^{V_j/V_T} - 1)$. Under the long-base condition $W_{p,n} \gg L_a^{(p,n)}$, the geometric factor $\coth(W/L_a) \to 1$; meanwhile, at low injection the ambipolar coefficients reduce to the minority-carrier values, $L_a^{(p)} \to L_n^{(p)}$ and $L_a^{(n)} \to L_p^{(n)}$, so the ICFs degenerate to $G_n \to qD_n^{(p)}/L_n^{(p)}$ and $G_p \to qD_p^{(n)}/L_p^{(n)}$. The extended equation thereby reduces systematically to the classical diode equation,

$$J(V_j) = \left(\frac{qD_n n_{p0}}{L_n} + \frac{qD_p p_{n0}}{L_p}\right)\left(e^{V_j/V_T} - 1\right) = J_0\left(e^{V_j/V_T} - 1\right), \tag{27}$$

with $J_0$ the classical saturation current. The classical result is thus contained as the low-injection, long-base limit of the extended equation; the regimes beyond this limit are the subject of the next subsection.

## D. Validity across injection level, geometry, and bias polarity

The *J*-$V_j$ characteristics are plotted in Fig.7 and are compared between the extended equation and the classical equation. The universality of the framework lies not in a single formula applying unchanged across all conditions, but in the self-consistent, physics-based mechanism that adapts the current law automatically as conditions change.

**(1) High injection.** All high-injection results follow directly from the exact boundary solution of Sec. III C. Under high injection, $\delta_0 \gg N$, the boundary excess becomes $\delta_0 \approx n_i e^{V_j/2V_T}$. For an asymmetric junction the lightly doped side enters high injection first; the crossover occurs when the injection term under the radical equals the doping term, $n_i^2 e^{V_j/V_T} = (N/2)^2$, which defines the *high-injection onset junction voltage* [11,12]

$$V_j^* \equiv 2V_T \ln\frac{N}{2n_i}, \qquad N = \min(N_A, N_D), \tag{28}$$

a quantity fixed solely by material and doping, independent of base geometry, always below $V_0$, delimiting the high-injection window $[V_j^*, V_0)$. Within that window the heavily doped side remains in low injection, so for a p$^+$–n junction

$$J \approx G_p\, \delta_0^{(n)} = \frac{qD_p^{(n)}}{L_a^{(n)}} \coth\frac{W_n}{L_a^{(n)}}\, n_i e^{V_j/2V_T} \qquad \text{(high injection, p}^+\text{–n)}, \tag{29}$$

while for symmetric doping both bases enter high injection simultaneously and the two ICF terms add. The net current thus grows with ideality factor $\mathrm{n} = 2$ while retaining the base-width dependence through the coth factor: the high-injection roll-off follows from first principles without an external series resistance [10–12]. The crossover is gradual and is quantified by the *effective ideality factor*: because the net current is proportional to the boundary excess in every regime, differentiating the quadratic boundary relation gives the closed form

$$\mathrm{n}(V_j) \equiv \left[\frac{d\ln J}{d(V_j/V_T)}\right]^{-1} = \frac{d(V_j/V_T)}{d\ln\delta_0} = \frac{2\delta_0(V_j) + N}{\delta_0(V_j) + N}, \tag{30}$$

which evolves continuously from 1 ($\delta_0 \ll N$) through exactly 3/2 at $\delta_0 = N$ to 2 ($\delta_0 \gg N$).

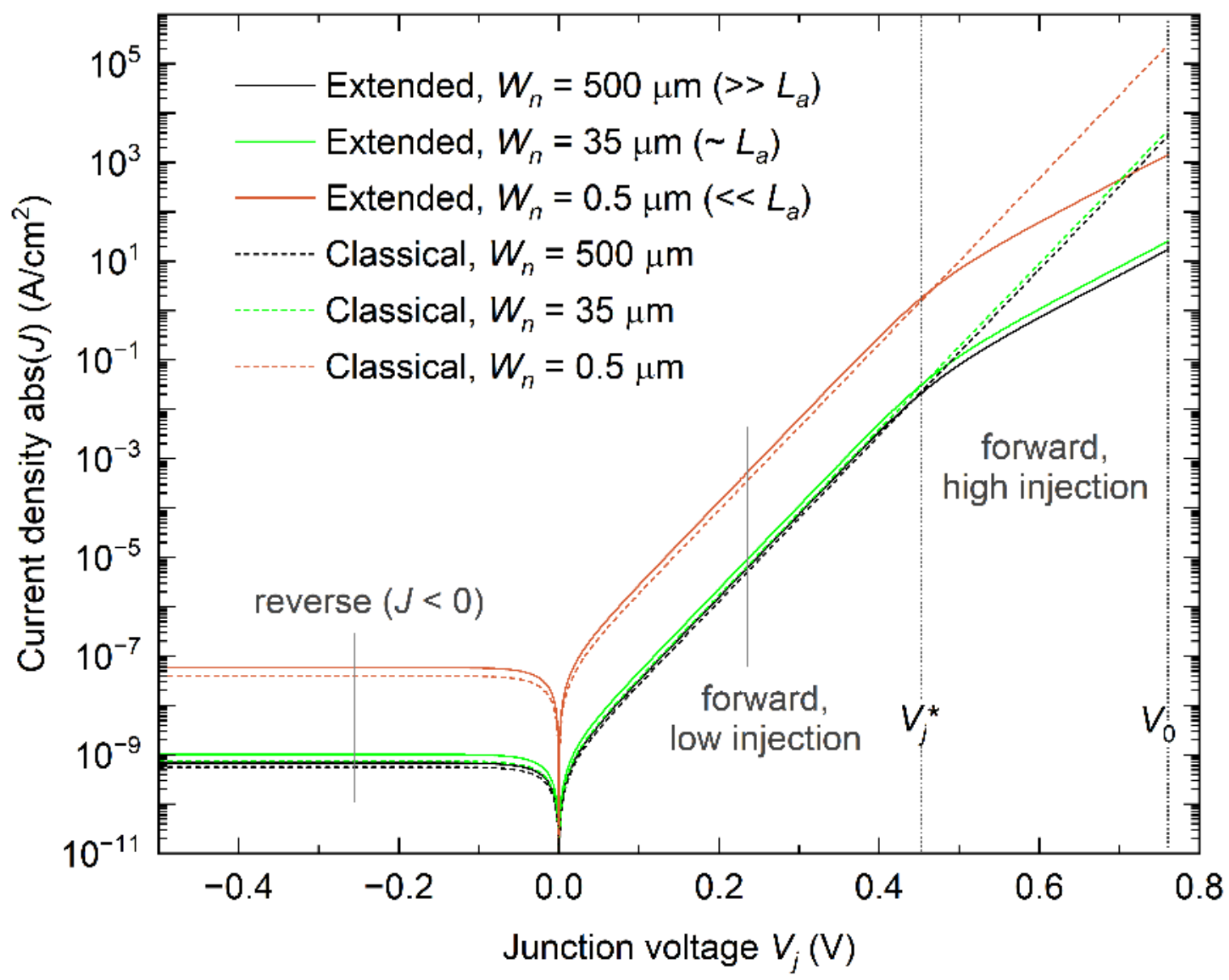

**FIG. 7.** Current density $J$ versus junction voltage $V_j$ predicted by the extended diode equation (solid) and by the classical diode equation (dashed) for the asymmetric p$^+$–n junction, over three n-base widths spanning the long-base ($W = 500\ \mu$m), intermediate ($W = 35\ \mu$m), and short-base ($W = 0.5\ \mu$m) regimes. The extended law combines the exact boundary excesses with the injection conversion factors $G = qD/L_a \coth(W/L_a)$; the classical curves use $J = J_0[\exp(V_j/V_T) - 1]$ with the geometry-dependent $J_0$. The vertical lines mark the high-injection onset $V_j^* = 2V_T \ln(N_D/2n_i) \approx 0.44$ V and the built-in potential $V_0 \approx 0.76$ V. Below $V_j^*$ the two characteristics coincide exactly (n = 1). Above $V_j^*$ the extended characteristics roll off toward $J \propto \exp(V_j/2V_T)$ (n → 2) from first principles and remain finite as $V_j \to V_0$, whereas the classical exponential diverges; the separation reaches a factor of ~ 200 at $V_j = 0.75$ V, identically for all three base widths. The base-width dependence enters only through $\coth(W/L_a)$: the long- and intermediate-base curves nearly overlap (coth ≈ 1 and 1.32), whereas the short-base current is enhanced by $\coth(W/L_a^{(n)}) \approx 42.8$ uniformly over the entire bias range.

**(2) Long base at arbitrary injection.** With $W_{p,n} \gg L_a^{(p,n)}$, the profile reduces to the exponential form and the current becomes width-independent,

$$J = \frac{qD_n^{(p)}\delta_0^{(p)}}{L_a^{(p)}} + \frac{qD_p^{(n)}\delta_0^{(n)}}{L_a^{(n)}} \qquad \text{(long base, arbitrary injection)}, \tag{31}$$

whose low-injection limit is precisely the classical equation of Sec. VI C.

**(3) Short base.** With $W_{p,n} \ll L_a^{(p,n)}$, the ICFs simplify to $G_n \to qD_n^{(p)}/W_p$ and $G_p \to qD_p^{(n)}/W_n$, and the current becomes explicitly width-dependent,

$$J = \frac{qD_n^{(p)}\delta_0^{(p)}}{W_p} + \frac{qD_p^{(n)}\delta_0^{(n)}}{W_n} \qquad \text{(short base, arbitrary injection)}. \tag{32}$$

Minority carriers traverse the base before recombining; the gradient is steeper and the current larger than in the long-base case at the same bias.

**(4) Reverse bias and reverse saturation.** As $V_j(< 0) \to -\infty$, the boundary excesses saturate at $\delta_0^{(n)} \to -p_{n0}$ and $\delta_0^{(p)} \to -n_{p0}$, and the current saturates at

$$J = -(G_n n_{p0} + G_p p_{n0}) = -J_0' \qquad \text{(reverse saturation)}, \tag{33}$$

where $J_0'$ differs slightly from the classical $J_0$ through the ICFs, which retain the effects of finite base width and non-negligible drift.

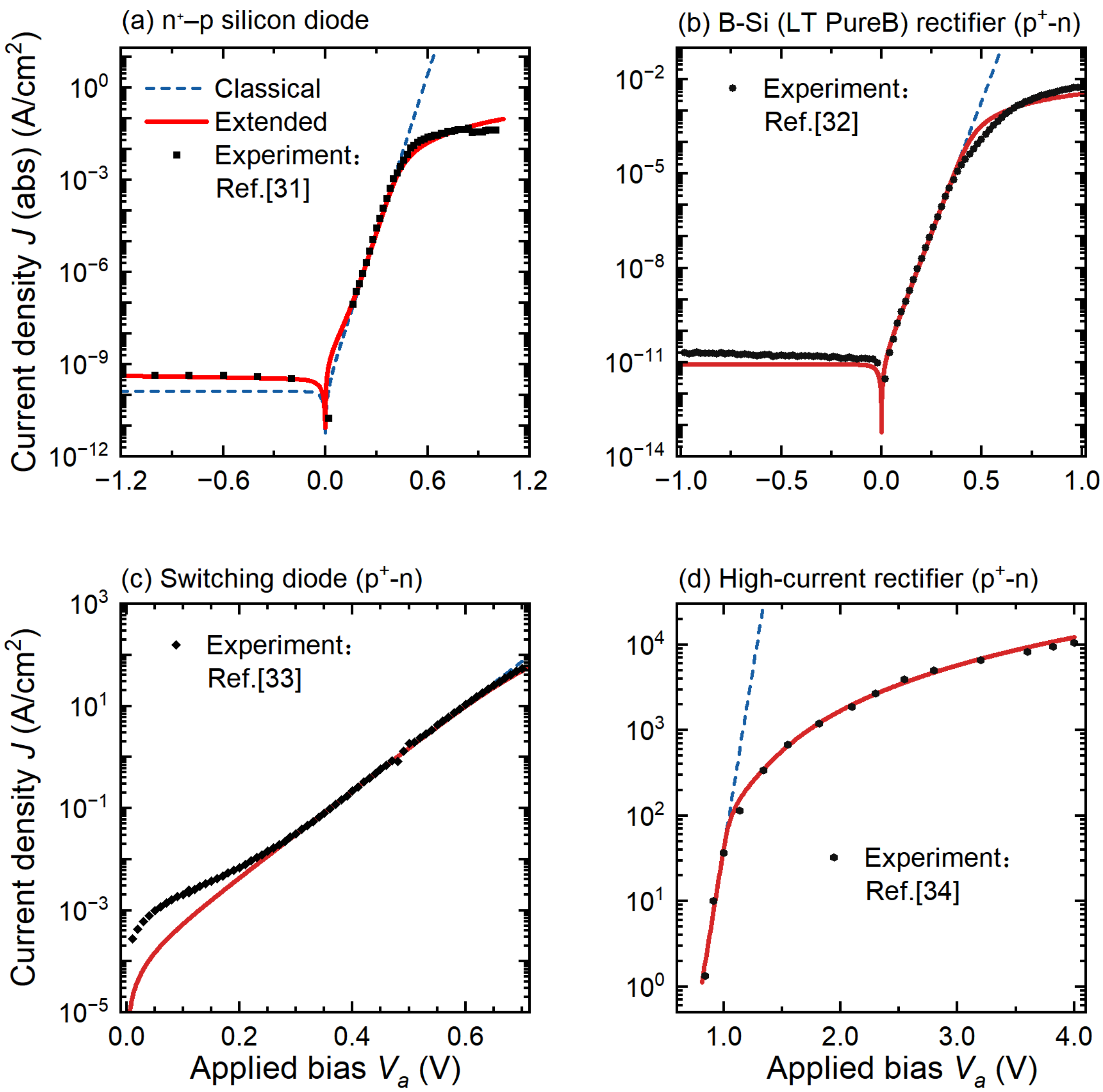

**FIG. 8.** Extended diode equation (solid red) compared with published experimental characteristics (symbols) of four reported devices, with the classical equation (dashed blue) shown for reference. The experimental points are digitized from the published figures; the samples are selected for their comparatively complete structural information; the model parameters are taken directly from the original reports or back-inferred from the information stated there; and each calculation is performed at the measurement temperature reported in the source (a) $n^+$–p silicon diode, measured at 20 °C [32]: the forward and reverse branches are both reproduced over the full ±1.2 V range, and the extended law follows the data through the high-injection roll-off that the classical exponential misses. (b) B-Si (LT PureB) rectifier [33] (represented by effective junction parameters): agreement over nearly nine decades of current. (c) Commercial switching diode (1N4148), measured at 47.8 °C [34]: agreement from $10^{-5}$ to $10^2$ A/cm². (d) High-current planar p–n rectifier (pulsed measurement) [35]: the deep roll-off between 1 and 4 V is reproduced, while the classical law overestimates the current by up to two orders of magnitude. No parameter is adjusted to improve the agreement, and no empirical series resistance is appended.

The same law is confronted with experiment in Fig. 8: the measured forward and reverse characteristics of four independently reported diodes [32–35], spanning both junction polarities and injection levels from leakage to deep high injection, are overlaid with the extended and classical predictions. The extended curves reproduce the reverse plateau, the ideal exponential region, and the high-injection roll-off in each case, using only the documented device parameters, whereas the classical law departs once high injection is entered.

## E. Testable predictions

Three parameter-free, experimentally testable predictions follow from the extended diode equation as predicted in Fig. 9: (i) the *universal ideality-factor transit* $\mathrm{n} = (2\delta_0 + N)/(\delta_0 + N)$ (Fig. 9a), which passes through exactly 3/2 at $\delta_0 = N$ regardless of material, doping, or geometry; (ii) the *high-injection onset voltage* $V_j^* = 2V_T \ln(N/2n_i)$ (Fig.9b), which scales with doping as $2V_T \ln 10 \approx 119.3$ mV per decade of $N$ at room temperature; and (iii) the *geometry factor* $\coth(W/L_a)$, which survives into high injection and enhances the short-base current uniformly (by a factor of ~70 for $W_n = 0.5$ $\mu$m, as shown in Fig.7), in contrast to the textbook expectation that high-injection behavior is geometry-independent. The experimental verification route for all three is discussed in Sec. IX D.

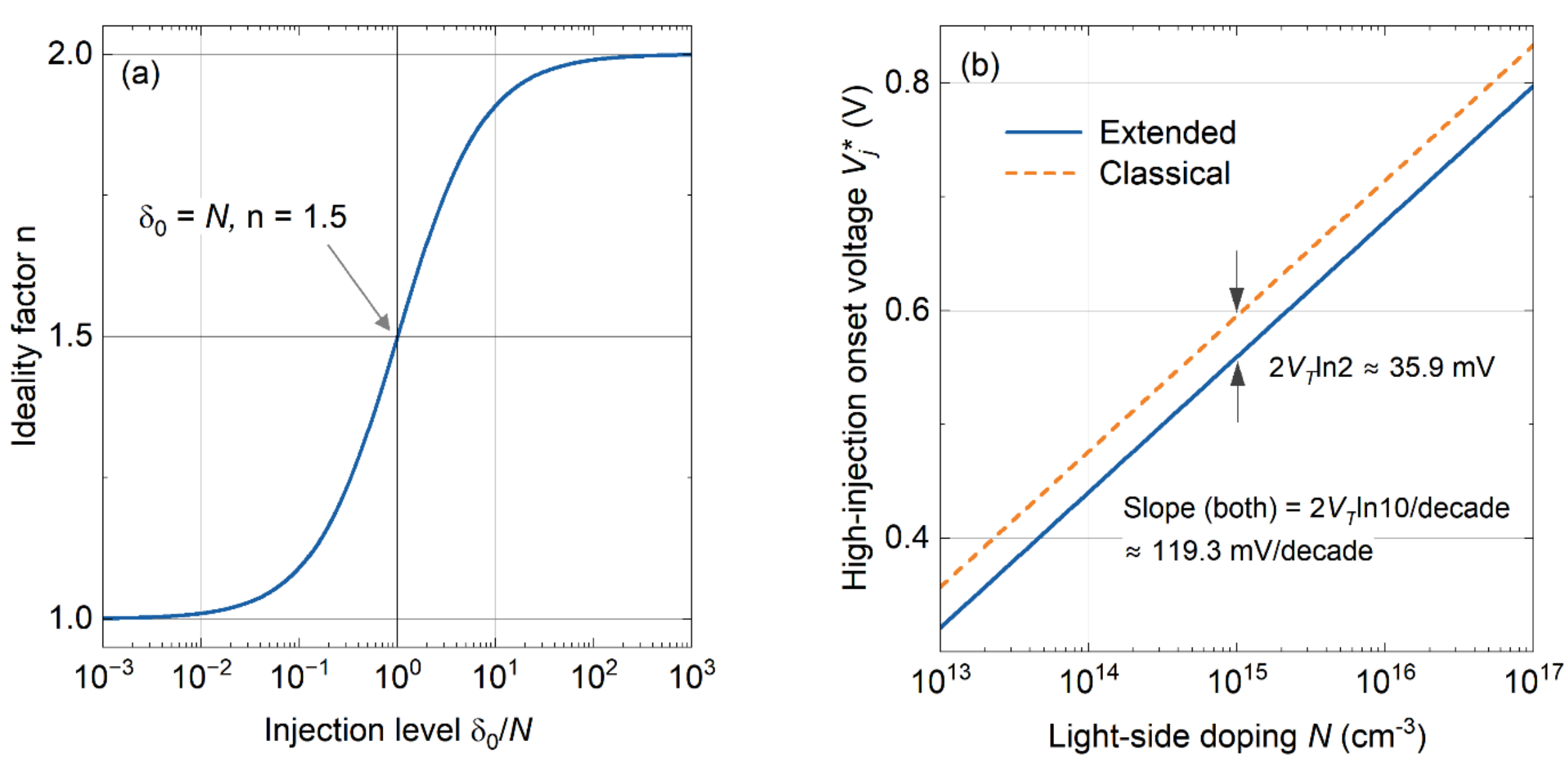

**FIG. 9.** Two of the three testable predictions of the extended diode equation (the third is shown in Fig. 7). (a) Effective ideality factor versus normalized boundary excess: n transits universally from 1 to 2, passing through exactly 3/2 at $\delta_0 = N$. (b) High-injection onset voltage versus the lighter doping level, compared between extended (blue) and classical (orange), both scales as 119 mV/decade of doping at 300 K.

# VII. Power dissipation

The two theories also differ in what the device power budget looks like. For a two-terminal device carrying current density $J$ under applied bias $V_a$, the dissipated power density is always $P = JV_a$; the difference lies in what $V_a$ means and in how $P$ is partitioned.

**Classical theory.** The base is field-free, so $V_j = V_a$; the current is $J = J_0(e^{V_a/V_T} - 1)$, and the entire dissipation is ascribed to the junction:

$$P^{\text{cls}}(V_a) = JV_a = J_0\left(e^{V_a/V_T} - 1\right)V_a \equiv P_j. \tag{34}$$

There is no base voltage drop, hence no base dissipation, by construction.

**Extended theory.** The self-consistent base field produces finite base drops (Sec. IV C), and the power separates exactly into a junction part and two base parts:

$$P^{\text{ext}}(V_a) = JV_a = J(V_j + V_{\text{ohmic}} + V_{\text{Dember}}) = P_j + P_{\text{ohmic}} + P_{\text{Dember}}. \tag{35}$$

Here $P_j$ is the power released by carriers crossing the junction; $P_{\text{ohmic}} = \int J^2/\sigma(x)\,dx \geq 0$ is genuine Joule heating of the quasi-neutral base, and $P_{\text{Dember}}$ is the work done against the concentration-gradient-induced field. Energy conservation $JV_a = P_j + P_{\text{ohmic}} + P_{\text{Dember}}$ is satisfied identically.

A second difference appears already at the junction itself. At fixed current (hence fixed boundary excess $\delta_0$), the exact boundary relation $\delta_0(\delta_0 + N_D) = n_i^2 e^{V_j/V_T}$ requires a larger junction voltage than the classical Boltzmann form:

$$\Delta V_j \equiv V_j^{\text{ext}} - V_j^{\text{cls}} = V_T \ln\left[\frac{(\delta_0 + N_D)\,\delta_0}{(\delta_0 + p_{n0})\,N_D}\right] \approx V_T \ln\left(1 + \frac{\delta_0}{N_D}\right) \geq 0, \tag{36}$$

because the junction voltage must sustain the excess *majority* population as well. The classical estimate of the power needed to maintain a given current is therefore too low even before the base drops are counted.

**Fixed-bias comparison.** At fixed $V_a$, the classical theory makes two compounding errors: it takes the full bias as the junction voltage, and it ignores the high-injection roll-off. Both inflate the current, so

$$\frac{P^{\mathrm{cls}}}{P^{\mathrm{ext}}} = \frac{J^{\mathrm{cls}}(V_a)}{J^{\mathrm{ext}}(V_a)} \geq 1. \tag{37}$$

The ratio at fixed $V_a$ is larger than the current ratio of $\sim 200 \times$ at fixed $V_j$ (Fig. 7), because part of the applied bias is consumed by the base drops (Fig. 5) and the classical error is compounded as shown in Fig. 10a: for the long-base device, $P^{\mathrm{cls}}/P^{\mathrm{ext}}$ rises from 6 at $V_a = 0.5$ V through $\sim 10^2$ at 0.6 V to $\sim 1.2 \times 10^4$ at 0.75 V; for the short-base device, from $\sim 24$ at 0.6 V to $\sim 3 \times 10^3$ at 0.8 V. The classical theory is thus over-optimistic about efficiency and over-pessimistic about the thermal load at a given voltage.

**Fixed-current comparison.** At fixed $J$, the actual device requires $V_a = V_j^{\mathrm{ext}} + V_{\mathrm{ohmic}} + V_{\mathrm{Dember}}$, whereas the classical theory believes the terminal voltage to be $V_j^{\mathrm{cls}}$. The classical estimate therefore *understates* the dissipation by

$$\Delta P = P^{\mathrm{ext}} - P^{\mathrm{cls}} = J(V_{\mathrm{ohmic}} + V_{\mathrm{Dember}}) + JV_T \ln\left(1 + \frac{\delta_0}{N_D}\right) \geq 0, \tag{38}$$

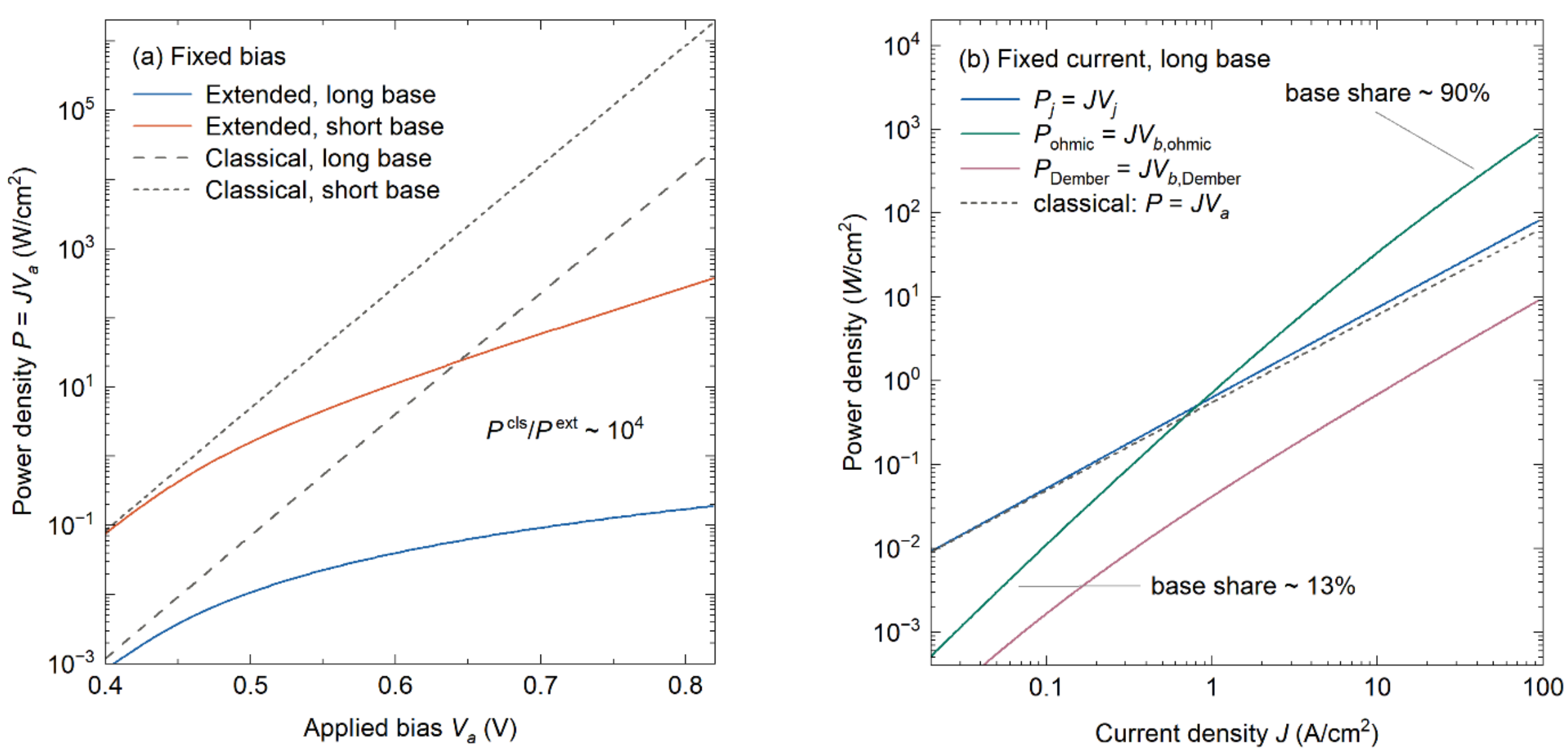

**FIG. 10.** Power dissipation of the asymmetric p$^+$–n junction, extended versus classical theory. (a) Fixed applied bias: $P = JV_a$ for the long-base ($W = 300$ $\mu$m) and short-base ($W = 0.5$ $\mu$m) devices. The extended curves (solid) incorporate the voltage partition and the high-injection roll-off; the classical curves (gray) take $V_j = V_a$ and the unbounded exponential, overestimating the power by $6 \times$ at 0.5 V rising to $\sim 10^4 \times$ at 0.75 V (long base) (b) Fixed current, long base: the exact partition $P = JV_j + JV_{\mathrm{ohmic}} + JV_{\mathrm{Dember}}$. The classical estimate (dashed) assigns everything to the junction and underestimates the total increasingly with current; the base (ohmic plus Dember) share rises from 13% to 90% over the range shown. Parameters as in Fig. 5.

i.e., it misses the base Joule heating, the Dember work, and the extra junction voltage that sustains the excess majority carriers. Since $V_{\text{ohmic}}$ grows super-linearly with $J$ in a lightly doped base, the underestimate grows rapidly with current: from 20% at $J = 0.05$ A/cm$^2$ to a factor of $\sim 4$ at 4.7 A/cm$^2$ in the long-base device, and 7–36% over the same range in the short-base device. The classical theory is over-optimistic about the voltage cost of sustaining a given current, precisely the quantity that matters in power-device operation [8,9].

**Power partition.** The most consequential result concerns where the energy actually goes. As shown in Fig. 10b and Table 1, at high current in a lightly doped long base, the dominant share of the dissipation migrates from the junction into the quasi-neutral base as Joule heat: the base share rises from 13% at $J = 0.05$ A/cm$^2$ to 75–90% at $J \gtrsim 5$ A/cm$^2$. The Dember contribution is small (a few percent of the base power) but is a genuinely new term with no classical counterpart. Thermal design based on the classical "all heat at the junction" picture misplaces the heat source, whereas the extended framework places it correctly through the exact partition.

**Table 1.** Power partition in the long-base p$^+$-n device ($W_n = 300$ $\mu$m, extended theory). Power density in W/cm$^2$.

| $J$ (A/cm$^2$) | $P_j$ | $P_{\text{ohmic}}$ | $P_{\text{Dember}}$ | Base share |
|---|---|---|---|---|
| 0.05 | 0.0247 | 0.0031 | 0.0005 | 13% |
| 0.47 | 0.281 | 0.191 | 0.016 | 42% |
| 4.7 | 3.32 | 9.79 | 0.28 | 75% |
| 46 | 38.7 | 327 | 4.1 | 90% |

# VIII. Minority-carrier velocity and base transit time

## A. Definitions and the classical limits

We consider the n-type base of the asymmetric p$^+$-n junction, with holes as the minority carriers, $n(x) = N_D + \delta(x)$ and $p(x) = p_{n0} + \delta(x)$. The local average velocity of the minority carriers, the base transit time, and the charge-control time are defined as

$$v_p(x) \equiv \frac{J_p(x)}{q\, p(x)}, \tag{39}$$

$$\tau_t \equiv \int_0^W \frac{dx}{v_p(x)}, \tag{40}$$

$$\tau_Q \equiv \frac{Q_p}{J_p(0)}, \quad Q_p = q\int_0^W \delta\,(x)\,dx, \tag{41}$$

the last being an alternative, device-level measure (stored charge per unit injected current) [36]. In the classical theory, $E = 0$ a priori and the motion is purely diffusive, $v_p(x) = -(D_p/p)\,d\delta/dx$; the textbook results follow [37,38]: for the long base, $v_p = D_p/L_p$ (position-independent), $\tau_t = \tau_p(W/L_p - \ln 2) \approx (W/L_p)\tau_p$, and $\tau_Q = \tau_p$; for the short base, $v_p(x) = D_p/(W - x)$ and $\tau_t = \tau_Q = W^2/2D_p$.

## B. The extended drift–diffusion velocity

With the self-consistent base field of Sec. IV A retained, the minority-carrier velocity becomes

$$v_p(x) = \mu_p E(x) - \frac{D_p}{p}\frac{d\delta}{dx} = \frac{\mu_p J}{q(\mu_n n + \mu_p p)} - \left[\frac{\mu_p(D_n - D_p)}{\mu_n n + \mu_p p} + \frac{D_p}{p}\right]\frac{d\delta}{dx}. \tag{42}$$

In the low-injection limit ($\delta \ll N_D$) the field terms are perturbations of order $\delta_0/N_D$, and the classical result is recovered, as required by the correspondence principle. In the high-injection limit ($n \approx p \approx \delta \gg N_D$), the expression closes into

$$v_p(x) = \frac{\mu_p}{\mu_n + \mu_p}\frac{J}{q\,\delta(x)} - D_a\frac{1}{\delta}\frac{d\delta}{dx}, \qquad D_a = \frac{2D_n D_p}{D_n + D_p}, \tag{43}$$

the diffusive part is upgraded from $D_p$ to the two-carrier ambipolar value $D_a$ (the internal field accelerates the slower species), and a genuine drift part appears. For the long base at high injection, $\delta(x) = \delta_0 e^{-x/L_a}$ and $J = q(D_p/L_a)\delta_0$, giving explicitly

$$v_p(x) = \frac{\mu_p}{\mu_n + \mu_p}\frac{D_p}{L_a}e^{\frac{x}{L_a}} + \frac{D_a}{L_a}. \tag{44}$$

The drift term grows exponentially along the base, so the carriers accelerate toward the ohmic contact; the second term raises the diffusion velocity by $\sqrt{D_a/D_p}$ relative to the classical value.

## C. Numerical results

The two theories agree to within 1% at low injection ($\delta_0 = 10^{12}$ cm$^{-3}$ $= 0.01N_D$) for all base widths, confirming the correspondence. At high injection ($\delta_0 = 10^{16}$ cm$^{-3}$ $= 100N_D$; $D_a = 18.3$ cm$^2$/s, $L_a = 42.7$ $\mu$m), the extended theory predicts systematically faster minority carriers, as shown in Fig. 11 and Table 2: the boundary velocity is enhanced by 43–73%, and the base transit time is shortened by ~40% for the shorter bases and by a factor of $\sim 6$ for the long base, where the drift term grows along $x$ and dominates the traversal. The field that the classical theory discards is what sweeps the minority carriers across the base at high current levels; this is the transport-level origin of the speed advantage of junction devices operated in high-injection regime, and it is captured only by the extended framework.

The charge-control measure behaves differently: at high injection $\tau_Q^{\text{ext}} = 1.47$ $\mu$s exceeds $\tau_Q^{\text{cls}} =$ $1.0$ $\mu$s for the long base (more stored charge through $L_a > L_p$, combined with the n $= 2$ current roll-off), while for the short base both theories give $\tau_Q = W^2/2D_p$ identically, because the ICF reduces to $qD_p/W_n$ in both frameworks. The two time measures $\tau_t$ and $\tau_Q$ therefore answer different questions; their relation to device-level modeling is taken up in Sec. IX C.

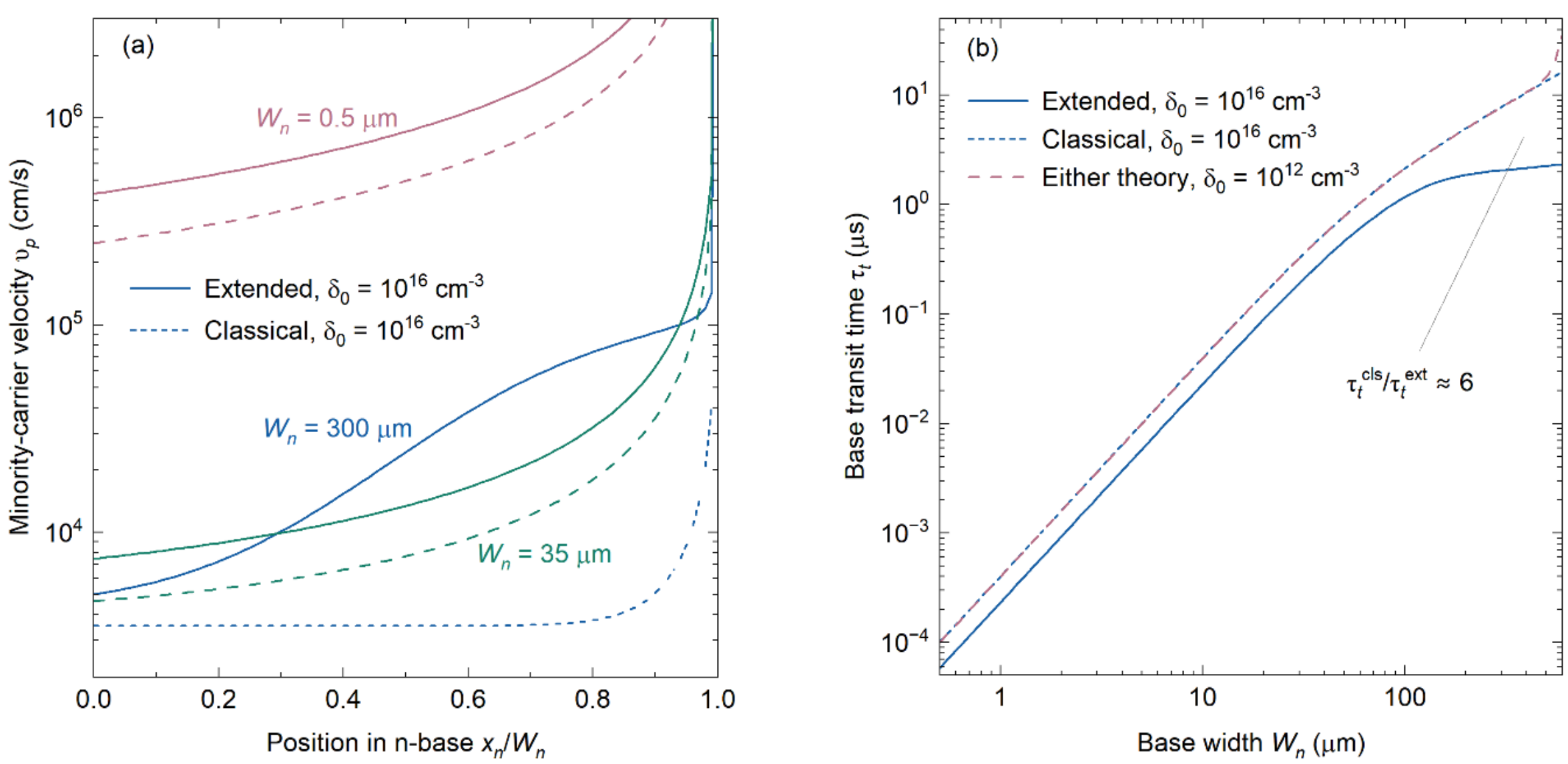

**FIG. 11.** Minority-carrier velocity and base transit time, extended (drift–diffusion) versus classical (diffusion-only) theory, n base of the asymmetric p$^+$–n junction. (a) Local velocity $v_p(x)$ along the base at high injection ($\delta_0 = 10^{16}$ cm$^{-3}$) for base widths $W = 300$, 35, and 0.5 $\mu$m. The extended curves (solid) exceed the classical ones (dashed) throughout: the ambipolar upgrade raises the diffusion component by $\sqrt{D_a/D_p} \approx 1.21$, and the drift component grows along the base, steeply near the end contact where the conductivity is unmodulated. (b) Base transit time versus base width. At high injection the extended transit time (blue) falls below the classical one (gray dashed) increasingly with $W$, reaching a factor of $\sim 6$ at $W = 500$ $\mu$m; at low injection ($\delta_0 = 10^{12}$ cm$^{-3}$, dotted) the two theories coincide, as they must. Parameters as in Fig. 5.

**Table 2.** Minority-carrier boundary velocity and base transit time at high injection ($\delta_0 = 10^{16}$ cm$^{-3}$), extended versus classical theory. Quantities are evaluated with high-injection conductivity (σ = 2.95 S/cm) at $x = x_{n0}$ for the QNR-n base. Velocities in cm/s, times in s, $E$ in V/cm.

| $W_n$ (μm) | $E$ (V/cm) | $\upsilon_p^{ext}$ (cm/s) | $\upsilon_p^{cls}$ (cm/s) | Ratio ($\upsilon_p^{ext}/\upsilon_p^{cls}$) | $\tau_t^{ext}$ (s) | $\tau_t^{cls}$ (s) | Ratio ($\tau_t^{ext}/\tau_t^{cls}$) |
|---|---|---|---|---|---|---|---|
| 500 | 4.43 | $5.03 \times 10^3$ | $3.52 \times 10^3$ | 1.43 | $2.26 \times 10^{-6}$ | $1.35 \times 10^{-5}$ | 0.17 |
| 35 | 6.57 | $7.45 \times 10^3$ | $4.64 \times 10^3$ | 1.61 | $2.53 \times 10^{-7}$ | $4.29 \times 10^{-7}$ | 0.59 |
| 0.5 | 378.6 | $4.30 \times 10^5$ | $2.48 \times 10^5$ | 1.73 | $5.8 \times 10^{-11}$ | $1.01 \times 10^{-10}$ | 0.58 |

# IX. Discussion

## A. Relation to earlier high-injection treatments

The failure of the low-injection theory at high current has been addressed before, but always case by case. Hall's early account of power rectifiers [10], Fletcher's analysis of high-current junction devices [11], and Herlet's treatment of silicon power rectifiers at high current densities [12,39] each describe particular regimes: conductivity modulation of a lightly doped region, the approach to an n = 2 characteristic, double-injection behavior [13,40], through regime-specific approximations. Textbooks incorporate the same physics phenomenologically, as "conductivity modulation" [3,8], while the ambipolar-transport foundations go back to van Roosbroeck [23,24]. The present framework provides a unifying closure rather than another regime-specific treatment: the exact two-sided boundary relation (Sec. III C) converts the high-injection crossover into an algebraic property of the junction law, so that the n = 1 → 2 transit, its onset voltage, and its geometry dependence all follow without regime-specific assumptions; and the base drops, previously lumped into an empirical series resistance, are obtained exactly from the self-consistent field (Sec. IV C). The earlier results are recovered as regime limits of the extended equation (Sec. VI D).

## B. Validity and limitations

The framework's validity boundaries are explicit and, where a correction exists, quantifiable. (i) *SCR generation–recombination* is neglected in the closure (Sec. VI A), as in the classical theory;

in real Si diodes the Sah–Noyce–Shockley component [31] adds an $\mathrm{n} \approx 2$ contribution at low bias, which must be subtracted before the present $\mathrm{n} = 1 \rightarrow 2$ transit can be read off measured characteristics. (ii) *Frozen ambipolar coefficients*: the closed form pairs the exact, injection-dependent boundary excess with transport coefficients evaluated at their low-injection limits. Retaining the injection dependence of $D_a$ and $L_a$ would forfeit the closed-form solution and necessitate numerical treatment; the justification parallels the familiar use of the Einstein relation, derived under equilibrium yet valid for moderate departures from it [3,6]. The ambipolar coefficients vary only by a bounded factor of order unity between the low- and high-injection limits (in the n base, from $D_p$ to $2D_nD_p/(D_n + D_p)$), so the result is asymptotically exact in the boundary injection level and first-order accurate in the transport coefficients. (iii) *Neglected ambipolar drift*: the perturbative elimination of $\mu_a E\, d\delta/dx$ (Sec. III A) is bounded in closed form: the associated ambipolar Péclet number satisfies $|\mathrm{Pe}_a| \leq 1/4(b + 1) \approx 6\%$ for Si, peaking at $\delta_0 \approx 0.75\, n_0$ and vanishing in both injection limits (Appendix A); at high injection the hyperbolic profile becomes asymptotically exact, since $\mu_a \rightarrow 0$ regardless of the field magnitude. (iv) *Constant material parameters*: bias-independent $\mu$ and $\tau$ are used throughout; at the highest injections, mobility degradation, Auger recombination, and self-heating (neglected here) would increase the base drops, so the ohmic drop reported in Sec. VII is a lower bound to what a fully coupled electro-thermal model would give [8,9,41]. (v) *Structure idealization*: uniformly doped, abrupt, one-dimensional junctions are assumed; graded bases, heterojunctions, and two-dimensional effects are outside the present scope, as are organic and disordered semiconductors, for which the dielectric-relaxation bound of Sec. II would have to be re-examined.

## C. Implications for device modeling

Two consequences follow for device engineering. First, the empirical series resistance of compact diode models [42,43] is replaced here by a physics-based partition: the difference between terminal bias and junction voltage is the exact base drop of Sec. IV C, computable from doping, geometry, and lifetime, rather than an adjustable resistor. This offers a route to extend charge-control-based compact models [36,44] into the high-injection regime without free parameters. Second, the two time measures of Sec. VIII must not be conflated: the transit time $\tau_t$, shortened by the base field at high injection, governs the speed of traversal, whereas the charge-control time $\tau_Q$ can *increase* (more stored charge at $L_a > L_p$). Device-level speed measures that equate the two will misjudge high-injection operation in opposite directions depending on which is used. Finally, the power-partition result of Sec. VII relocates the dominant heat source from the junction into the base at high current; thermal models of power diodes that place the dissipation at the junction [8,9] misestimate both the temperature profile and the efficiency budget.

### D. Experimental testability

The three predictions of Sec. VI E are designed to be checkable against existing device data, without dedicated new experiments. (i) The ideality-factor transit $\mathrm{n} = (2\delta_0 + N)/(\delta_0 + N)$, in particular the exact value 3/2 at $\delta_0 = N$, is parameter-free: it can be extracted from the forward characteristics of any diode whose doping, base width, and lifetime are independently known, with $\delta_0$ obtained from the measured current through the ICF, so that the transit, unlike the current magnitude, involves no adjustable parameters. The classic forward characteristics of silicon power rectifiers measured by Herlet [12] are an ideal data source: the structures are fully characterized, the high-injection regime is deeply entered, and digitization of the published curves suffices to test the transit quantitatively. (ii) The onset voltage $V_j^* = 2V_T \ln(N/2n_i)$ predicts a 119 mV/decade doping scaling, testable with a doping series of otherwise identical diodes (or across published datasets spanning different base dopings). (iii) The geometry factor $\coth(W/L_a)$ surviving into high injection is testable with an epitaxial-thickness series at fixed doping: the textbook expectation is that high-injection current becomes geometry-independent, whereas the extended law retains the full coth enhancement. Complementary transient measurements, such as reverse-recovery and open-circuit voltage decay [8,30,45], probe the same stored-charge physics and can cross-check the transit-time predictions of Sec. VIII. In each case the test would use third-party, independently characterized devices, which strengthens rather than weakens it.

## X. Conclusion

We have developed a self-consistent two-carrier drift–diffusion framework that restructures the theory of the p–n junction while preserving the classical results as demonstrable limits. The framework (i) establishes the excess majority carriers as electrostatically forced transport participants, bounded by the dielectric-relaxation ratio $\tau_d/\tau_p$; (ii) solves the ambipolar transport problem with exact two-sided boundary conditions, yielding a unified hyperbolic profile and boundary-to-boundary locking of the excess populations; (iii) obtains the self-consistent base field, separating exactly into ohmic and Dember components, and reinterprets the junction voltage as a self-limiting drop confined to the space-charge region, separated from the applied bias by exact base drops; (iv) exhibits all four current components in closed form, demonstrating current continuity plane by plane; (v) closes at device level into the extended diode equation $J = G_n\delta_0^{(p)} + G_p\delta_0^{(n)}$, valid across forward and reverse bias, low and high injection, and arbitrary base length, containing the Shockley equation as its low-injection, long-base limit and the high-injection n → 2 roll-off as an intrinsic consequence; and (vi) converts these results into device-level corrections that reverse the classical errors in both directions: the dissipated power is overestimated at fixed bias (by up to $\sim 10^4$) and underestimated at fixed current, with 75–90% of the dissipation migrating into the quasi-neutral base at high current, while the minority-carrier transit time is overestimated by up to a factor of six. Three parameter-free predictions: the universal ideality-factor transit through 3/2, the doping-set onset voltage $V_j^*$ scaling at 119 mV/decade, and the injection conversion factor surviving into high injection, are directly testable against existing device data. By eliminating the artificial separation between "minority-carrier physics" and "majority-carrier effects," and by deriving rather than assuming the key quantities, the framework achieves universality without sacrificing rigor. We expect this work to provide the physical basis for improved compact models of junction devices, and its extension to graded, heterostructure, and wide-bandgap junctions to be a natural continuation.

# Appendix A. Bound on the neglected ambipolar drift term

The elimination of $\mu_a E\, d\delta/dx$ from the ambipolar transport equation (Sec. III A) is justified as follows. Nondimensionalizing the full equation with $\xi = x/L_a$ and $L_a^2 = D_a\tau_a$ gives

$$\delta_{\xi\xi} + \varepsilon\,\delta_\xi - \delta = 0, \qquad \varepsilon \equiv \frac{\mu_a E L_a}{D_a},$$

where $\varepsilon$ plays the role of an ambipolar Péclet number, measuring drift against diffusion transport of the excess-carrier package. In the diffusion-dominated regime, $E \sim J/\sigma$ and $J \sim qD_a\delta_0/L_a$; substitution eliminates $D_a$ and $L_a$ exactly, leaving a pure concentration expression. For the n-type base with injection level $\chi \equiv \delta_0/n_0$, quasi-neutrality gives $n = n_0(1+\chi)$, $p \approx n_0\chi$, hence $p - n = -n_0$ and $bn + p = n_0[b + (b+1)\chi]$ with $b = \mu_n/\mu_p$; the doping cancels, leaving

$$|\varepsilon(\chi)| = \frac{b\,\chi}{[b + (b+1)\chi]^2}.$$

This vanishes in both limits: at low injection because the field is small, at high injection because $\mu_a \to 0$ as $p \to n$; it is largest only at intermediate injection. Maximizing over $\chi$ gives the peak position and value

$$\chi^* = \frac{b}{b+1}, \qquad \varepsilon_{\max} = \frac{1}{4(b+1)}.$$

For silicon, $b \approx 3$, so $\varepsilon_{\max} = 1/16 \approx 6\%$, occurring at $\delta_0 \approx 0.75\, n_0$, i.e., the intermediate-injection band; within the standard operating regimes the correction is far smaller. The neglected term is therefore not a divergence but a bounded, first-order-computable offset, capped over the entire injection range by a constant that depends only on the mobility ratio, and the hyperbolic profile is asymptotically *exact* at high injection, where $\mu_a \to 0$ irrespective of the field. Note also that the structural results of the framework (i.e., the current continuity, the voltage partition, and the retention of minority-carrier drift) do not depend on the profile shape at all: they follow from the continuity equations and the boundary constraints directly. Only the $\coth(W/L_a)$ relation between $\delta_0$ and $J$ carries the small intermediate-injection correction bounded above.

## Appendix B. Numerical methods and parameters

The numerical validation quoted in Sec. II was obtained in Ref. [15] by solving the full steady-state drift–diffusion system (the two continuity equations and Poisson's equation, without imposing quasi-neutrality) using the Scharfetter–Gummel discretization [46] within Gummel's decoupled iteration scheme [47], as described in standard treatments of semiconductor device simulation [48,49,50]. The independently tracked excess electron and hole profiles coincide everywhere in the quasi-neutral base, with the relative imbalance at the $\tau_d/\tau_p$ level, providing the numerical justification for imposing $\delta p(x) = \delta n(x)$ in the analytical framework.

All analytical evaluations in this paper use the single parameter set of Table 3 [51-53] (Si, 300 K; asymmetric $p^+$–n junction). The classical curves carry the same $\coth(W/L_p)$ geometry factor as the extended ones, so that every comparison isolates the framework difference rather than a geometry difference. The p base, being heavily doped and 1 $\mu$m wide, contributes negligibly to the base drops and power for this asymmetric junction; for symmetric or p-side-lightly-doped structures both bases must be counted.

**Table 3.** Parameter set used in all numerical evaluations (Si, 300 K) unless otherwise indicated.

| Quantity (unit) | n base | $p^+$ base |
|---|---|---|
| Doping ($cm^{-3}$) | $N_D = 1 \times 10^{14}$ | $N_A = 5 \times 10^{18}$ |
| $\mu_n$ ($cm^2$/V·s) | 1350 | 150 |
| $\mu_p$ ($cm^2$/V·s) | 480 | 50 |
| $D_n$ ($cm^2$/s) | 34.9 | 3.88 |
| $D_p$ ($cm^2$/s) | 12.4 | 1.29 |
| $\tau_n, \tau_p$ ($\mu$s) | 1.0 | 1.0 |
| $W_n, W_p$ ($\mu$m) | 500, 300, 35, 0.5 | 500, 300, 35, 0.5 |
| $\sigma_0$ (mS/cm) | 21.7 | $40.0 \times 10^3$ |
| $n_i$ ($cm^{-3}$) | $1.25 \times 10^{10}$ | |

## Acknowledgments

The authors thank Associate Professor Guipeng Liu for helpful suggestions, and Mr. Han Peng and Mr. Jidu Liu, master's students, for data processing and preparing some of the illustrations.

## References

[1] W. Shockley, "The theory of p–n junctions in semiconductors and p–n junction transistors," *Bell Syst. Tech. J.* **28**, 435–489 (1949).

[2] W. Shockley, *Electrons and Holes in Semiconductors* (Van Nostrand, New York, 1950).

[3] S. M. Sze and K. K. Ng, *Physics of Semiconductor Devices*, 3rd ed. (Wiley, Hoboken, 2007).

[4] R. F. Pierret, *Semiconductor Device Fundamentals* (Addison-Wesley, Reading, 1996).

[5] C. T. Sah, *Fundamentals of Solid-State Electronics* (World Scientific, Singapore, 1991).

[6] D. A. Neamen, *Semiconductor Physics and Devices: Basic Principles*, 4th ed. (McGraw-Hill, New York, 2012).

[7] A. S. Grove, *Physics and Technology of Semiconductor Devices* (Wiley, New York, 1967).

[8] B. J. Baliga, *Fundamentals of Power Semiconductor Devices* (Springer, New York, 2008).

[9] J. Lutz, H. Schlangenotto, U. Scheuermann, and R. De Doncker, *Semiconductor Power Devices: Physics, Characteristics, Reliability* (Springer, Berlin, 2011).

[10] R. N. Hall, "Power rectifiers and transistors," *Proc. IRE* **40**, 1512–1518 (1952).

[11] N. H. Fletcher, "The high current limit for semiconductor junction devices," *Proc. IRE* **45**, 862–872 (1957).

[12] A. Herlet, "The forward characteristic of silicon power rectifiers at high current densities," *Solid-State Electron.* **11**, 717–742 (1968).

[13] M. A. Lampert and P. Mark, *Current Injection in Solids* (Academic, New York, 1970).

[14] H. K. Henisch, *Rectifying Semi-Conductor Contacts* (Clarendon Press, Oxford, 1957).

[15] J. H. Yang, Y. Zhang, W. J. Chen, and G. P. Liu, "p–n junction theory in view of excess majority carriers," *EPL* **120**, 28004 (2017).

[16] B. G. Streetman and S. K. Banerjee, *Solid State Electronic Devices*, 7th ed. (Pearson, Upper Saddle River, NJ, 2015).

[17] R. S. Muller and T. I. Kamins, *Device Electronics for Integrated Circuits*, 3rd ed. (Wiley, New York, 2003).

[18] J. H. Yang, G. H. Sun, and T. L. Li, companion Letter: "The Extended Diode Equation" (to be submitted simultaneously).

[19] T. Kimoto and J. A. Cooper, *Fundamentals of Silicon Carbide Technology: Growth, Characterization, Devices, and Applications* (Wiley-IEEE, Singapore, 2014).

[20] H. Morkoç, *Handbook of Nitride Semiconductors and Devices*, Vol. 1 (Wiley-VCH, Weinheim, 2008).

[21] W. Shockley and W. T. Read, "Statistics of the recombinations of holes and electrons," *Phys. Rev.* **87**, 835–842 (1952).

[22] R. N. Hall, "Electron-hole recombination in germanium," *Phys. Rev.* **87**, 387 (1952).

[23] W. van Roosbroeck, “Theory of the flow of electrons and holes in germanium and other semiconductors,” *Bell Syst. Tech. J.* **29**, 560–607 (1950).

[24] W. van Roosbroeck, “The transport of added current carriers in a homogeneous semiconductor,” *Phys. Rev.* **91**, 282–289 (1953).

[25] R. A. Smith, *Semiconductors*, 2nd ed. (Cambridge University Press, Cambridge, 1978).

[26] J. P. McKelvey, *Solid State and Semiconductor Physics* (Harper & Row, New York, 1966).

[27] M. Lundstrom, *Fundamentals of Carrier Transport*, 2nd ed. (Cambridge University Press, Cambridge, 2000).

[28] H. Dember, “Über eine photoelektromotorische Kraft in Kupferoxydul-Kristallen,” *Phys. Z.* **32**, 554 (1931).

[29] M. A. Green, *Solar Cells: Operating Principles, Technology, and System Applications* (Prentice-Hall, Englewood Cliffs, 1982).

[30] S. K. Ghandhi, *Semiconductor Power Devices* (Wiley, New York, 1977).

[31] C. T. Sah, R. N. Noyce, and W. Shockley, “Carrier generation and recombination in p–n junctions and p–n junction characteristics,” *Proc. IRE* **45**, 1228–1243 (1957).

[32] E. Simoen, J. Vanhellemont, A. L. P. Rotondaro, and C. Claeys, “Static and low-frequency noise characteristics of $n^+$-p junction diodes fabricated in different silicon substrates,” Semicond. Sci. Technol. **10**, 1002–1008 (1995).

[33] V. Mohammadi, S. Nihtianov, and C. Fang, “A doping-less junction-formation mechanism between n-silicon and an atomically thin boron layer,” Sci. Rep. **7**, 13247 (2017).

[34] E. Cataldo, A. Di Lieto, F. Maccarrone, and G. Paffuti, “Measurements and analysis of current-voltage characteristic of a pn diode for an undergraduate physics laboratory,” arXiv:1608.05638 [physics.ed-ph] (2016).

[35] A. K. Jonscher, “Measurement of Voltage-Current Characteristics of Junction Diodes at High Forward Bias,” Journal of Electronics and Control **5**, 226–244 (1958).

[36] R. Beaufoy and J. J. Sparkes, "The junction transistor as a charge-controlled device," *ATE J.* **13**, 310–327 (1957).

[37] J. L. Moll, *Physics of Semiconductors* (McGraw-Hill, New York, 1964).

[38] Y. Taur and T. H. Ning, *Fundamentals of Modern VLSI Devices*, 2nd ed. (Cambridge University Press, Cambridge, 2009).

[39] E. Spenke, "Notes on the theory of the forward characteristic of power rectifiers," Solid-State Electron. **11**, 1119–1130 (1968).

[40] A. K. Jonscher, *Principles of Semiconductor Device Operation* (Wiley, New York, 1960).

[41] J. Dziewior and W. Schmid, "Auger coefficients for highly doped and highly excited silicon," Appl. Phys. Lett. **31**, 346–348 (1977).

[42] G. Massobrio and P. Antognetti, *Semiconductor Device Modeling with SPICE*, 2nd ed. (McGraw-Hill, New York, 1993).

[43] P. O. Lauritzen and C. L. Ma, "A simple diode model with reverse recovery," IEEE Trans. Power Electron. **6**, 188–191 (1991).

[44] H. K. Gummel and H. C. Poon, "An integral charge control model of bipolar transistors," *Bell Syst. Tech. J.* **49**, 827–852 (1970).

[45] D. K. Schroder, *Semiconductor Material and Device Characterization*, 3rd ed. (Wiley-IEEE Press, Hoboken, NJ, 2006).

[46] D. L. Scharfetter and H. K. Gummel, "Large-signal analysis of a silicon Read diode oscillator," *IEEE Trans. Electron Devices* **16**, 64–77 (1969).

[47] H. K. Gummel, "A self-consistent iterative scheme for one-dimensional steady state transistor calculations," *IEEE Trans. Electron Devices* **11**, 455–465 (1964).

[48] S. Selberherr, *Analysis and Simulation of Semiconductor Devices* (Springer, Vienna, 1984).

[49] D. Vasileska, S. M. Goodnick, and G. Klimeck, *Computational Electronics: Semiclassical and Quantum Device Modeling and Simulation* (CRC Press, Boca Raton, FL, 2010).

[50] R. W. Dutton and Z. Yu, *Technology CAD: Computer Simulation of IC Processes and Devices* (Kluwer Academic, Boston, 1993).

[51] N. D. Arora, J. R. Hauser, and D. J. Roulston, “Electron and hole mobilities in silicon as a function of concentration and temperature,” IEEE Trans. Electron Devices **29**, 292–295 (1982).

[52] D. B. M. Klaassen, “A unified mobility model for device simulation—I. Model equations and concentration dependence,” Solid-State Electron. **35**, 953–959 (1992).

[53] D. M. Caughey and R. E. Thomas, “Carrier mobilities in silicon empirically related to doping and field,” Proc. IEEE **55**, 2192–2193 (1967).

[54] H. Dember, “Über die Vorwärtsbewegung von Elektronen durch Licht,” Physikalische Zeitschrift **33**, 207 (1932).

[55] J. Tauc, “Generation of an emf in Semiconductors with Nonequilibrium Current Carrier Concentrations,” Rev. Mod. Phys. **29**, 308–324 (1957).

[56] M. Krčmar and W. M. Saslow, "Exact surface solutions for semiconductors: The Dember effect and partial currents," Phys. Rev. B **65**, 233313 (2002).